\documentclass[sn-nature]{sn-jnl}% Style for submissions to Nature Portfolio journals
\usepackage{graphicx}%
\usepackage{multirow}%
\usepackage{amsmath,amssymb,amsfonts}%
\usepackage{amsthm}%
\usepackage{mathrsfs}%
\usepackage[title]{appendix}%
\usepackage{xcolor}%
\usepackage{textcomp}%
\usepackage{manyfoot}%
\usepackage{booktabs}%
\usepackage{algorithm}%
\usepackage{algorithmicx}%
\usepackage{algpseudocode}%
\usepackage{listings}%
\usepackage{caption}
\usepackage{array}
\theoremstyle{thmstyleone}%
\theoremstyle{thmstyletwo}%

\theoremstyle{thmstylethree}%

\begin{document}

\title[A second-scale periodicity in an active repeating fast radio burst source]{A second-scale periodicity in an active repeating fast radio burst source}

%%=============================================================%%
%% GivenName    -> \fnm{Joergen W.}
%% Particle -> \spfx{van der} -> surname prefix
%% FamilyName   -> \sur{Ploeg}
%% Suffix   -> \sfx{IV}
%% \author*[1,2]{\fnm{Joergen W.} \spfx{van der} \sur{Ploeg}
%%  \sfx{IV}}\email{iauthor@gmail.com}
%%=============================================================%%

\author[1]{\fnm{Chen} \sur{Du}}

\author*[1,2]{\fnm{Yong-Feng} \sur{Huang}}\email{hyf@nju.edu.cn}

\author[3]{\fnm{Jin-Jun} \sur{Geng}}

\author[4,5]{\fnm{Hao-Xuan} \sur{Gao}}

\author*[6]{\fnm{Li} \sur{Zhang}}\email{lizhang.science@gmail.com}

\author[1]{\fnm{Chen} \sur{Deng}}

\author[7,8,9]{\fnm{Lang} \sur{Cui}}

\author[7,10]{\fnm{Jie} \sur{Liao}}

\author[7,2]{\fnm{Peng-Fei} \sur{Jiang}}

\author[11,6]{\fnm{Liang} \sur{Zhang}}

\author[12,13,8]{\fnm{Pei} \sur{Wang}}

\author[1]{\fnm{Chen-Ran} \sur{Hu}}

\author[1]{\fnm{Xiao-Fei} \sur{Dong}}

\author[14,1]{\fnm{Fan} \sur{Xu}}

\author[15,16,17]{\fnm{Liang} \sur{Li}}

\author[1]{\fnm{Ze-Cheng} \sur{Zou}}

\author[7,8,9]{\fnm{Abdusattar} \sur{Kurban}}

\affil[1]{\orgdiv{School of Astronomy and Space Science},
\orgname{Nanjing University}, \orgaddress{\city{Nanjing}
\postcode{210023}, \state{Jiangsu}, \country{China}}}

\affil[2]{\orgdiv{Key Laboratory of Modern Astronomy and
Astrophysics (Nanjing University)}, \orgname{Ministry of
Education}, \orgaddress{\city{Nanjing}  \postcode{210023},
\state{Jiangsu}, \country{China}}}

\affil[3]{\orgdiv{Purple Mountain Observatory},  \orgname{Chinese
Academy of Sciences}, \orgaddress{\city{Nanjing}
\postcode{210023}, \state{Jiangsu}, \country{China}}}

\affil[4]{\orgdiv{School of Physics and Astromoy}, \orgname{Anqing Normal University},
\orgaddress{\city{Anqing}  \postcode{246133}, \state{Anhui}, \country{China}}}

\affil[5]{\orgdiv{Institute of Astronomy and Astrophysics},  \orgname{Anqing Normal University},
\orgaddress{\city{Anqing}  \postcode{246133}, \state{Anhui}, \country{China}}}

\affil[6]{\orgdiv{Guizhou Provincial Laboratory of Big Data, College of Big Data and Information Engineering},  \orgname{Guizhou University},
\orgaddress{\city{Guiyang}  \postcode{550025}, \state{Guizhou},
\country{China}}}

\affil[7]{\orgdiv{Xinjiang Astronomical Observatory},
\orgname{Chinese Academy of Sciences}, \orgaddress{\city{Urumqi}
\postcode{830011}, \state{Xinjiang}, \country{China}}}

\affil[8]{\orgdiv{State Key Laboratory of Radio Astronomy and
Technology},  \orgname{Chinese Academy of Sciences},
\orgaddress{\city{Beijing}  \postcode{100101}, \country{China}}}

\affil[9]{\orgdiv{Xinjiang Key Laboratory of Radio Astrophysics},
\orgaddress{\city{Urumqi}  \postcode{830011}, \state{Xinjiang},
\country{China}}}

\affil[10]{\orgdiv{College of Astronomy and Space Science},
\orgname{University of Chinese Academy of Sciences},
\orgaddress{\city{Beijing}  \postcode{101408}, \country{China}}}

\affil[11]{\orgdiv{Guizhou Vocational College of Economics and
Business},  \orgaddress{\city{Duyun}  \postcode{558022},
\state{Guizhou}, \country{China}}}

\affil[12]{\orgdiv{CAS Key Laboratory of FAST, NAOC},
\orgname{Chinese Academy of Sciences}, \orgaddress{\city{Beijing}
\postcode{100101}, \country{China}}}

\affil[13]{\orgdiv{Institute for Frontiers in Astronomy and
Astrophysics},  \orgname{Beijing Normal University},
\orgaddress{\city{Beijing}  \postcode{102206}, \country{China}}}

\affil[14]{\orgdiv{Institute of Space Weather, School of Atmospheric Physics}, \orgname{Nanjing University of Information Science and Technology},  \orgaddress{\city{Nanjing}  \postcode{210044},
\state{Jiangsu}, \country{China}}}

\affil[15]{\orgdiv{Institute of Fundamental Physics and Quantum
Technology},  \orgname{Ningbo University},
\orgaddress{\city{Ningbo}  \postcode{315211}, \state{Zhejiang},
\country{China}}}

\affil[16]{\orgdiv{Department of Physics, School of Physical
Science and Technology},  \orgname{Ningbo University},
\orgaddress{\city{Ningbo}  \postcode{315211}, \state{Zhejiang},
\country{China}}}

\affil[17]{\orgdiv{INAF-Osservatorio Astronomico d'Abruzzo},
\orgaddress{\city{Teramo}  \postcode{64100}, \country{Italy}}}

%%==================================%%
%% Sample for unstructured abstract %%
%%==================================%%

\abstract{Fast radio bursts (FRBs) are fierce radio flashes from
the deep sky. Abundant observations have indicated that highly
magnetized neutron stars might be involved in these energetic
bursts, but the underlying trigger mechanism is still enigmatic.
Especially, the widely expected periodicity connected to the spin
of the central engine has never been discovered, which leads to
further debates on the nature of FRBs. Here we report the first
discovery of a $\sim$ 1.7 s period in the repeating source of FRB
20201124A. This is an active repeater, from which more than 2800
bursts have been detected over a total of 49 days. The phase-folding
method is adopted to analyze the bursts on each day separately.
While no significant periodic signal is found in most days, a clear
periodicity does appear on two specific days: a period of 1.706024(13) s
on MJD 59310, and a slightly larger period of 1.707968(9) s on MJD 59347.
A global Monte Carlo analysis based on all single-day datasets
yields a significance level of $5.5 \sigma$ for the periodicity.
A period derivative of $6.11(5)\times10^{-10}$ s s$^{-1}$ can be
derived from these two periods, which further implies a surface magnetic field
strength of $1.03\times10^{15}$ G and a spin-down age of $44$
years for the central engine. It is
concluded that FRB 20201124A should be associated with a young
magnetar.}

\maketitle

\section*{Introduction}
\label{sec:intro}

Fast radio bursts (FRBs) are radio transients with millisecond
duration and extremely high brightness temperature
\citep{lorimer2007Bright}. Over 800 FRB sources have been
discovered to date, with more than 60 classified as repeaters
\citep{Petroff2020,chime/frb2021First,xu2023Blinkverse}. FRBs
exhibit a diverse and perplexing phenomenology
\citep{pleunis2021Fast,zhang2023physics,hu2023Comprehensivea},
with the underlying physical mechanisms still largely unknown.
They may be connected with magnetars. Such a connection is
supported by the association of FRB 20200428 with an X-ray burst
from the galactic magnetar SGR 1935+2154
\citep{chime/frb2020Abright,mereghetti2020INTEGRAL,li2021HXMT}.
Consequently, models involving the magnetosphere of a magnetar,
analogous to the emission site of radio pulsars, have been widely
discussed. Evidence for such a magnetospheric origin is
accumulating. For instance, a recent study using scintillation to
constrain the size of the emission region of FRB 20221022A
\citep{Nimmo2025} reveals that it is small, supporting the
magnetosphere origin. Furthermore, the S-shaped swing in the
polarization position angle observed in a burst from the same
source is a clear signature of emission from a rotating
magnetosphere \citep{Mckinven2025}.

A natural expectation from the magnetospheric origin models is
that FRBs' times of arrival (TOAs) would show periodicity tied to
the magnetar rotation. Currently, some FRBs have been observed to
exhibit long-timescale periodicity, such as FRB 20180916B's
16.35-day period with a 5-day active window
\citep{chime/frb2020Periodic,pastor-marazuela2021Chromatic,gopinath2023Propagationa},
FRB 20121102A's $\sim$ 165-day period with a duty cycle exceeding
50\% \citep{rajwade2020Possible}, and FRB 20240209A's putative
$\sim$ 126-day period \citep{Pal2025}. Usually, such long periods
are unlikely related to the rotation of magnetars, as long period
pulsars are generally old and should be radio-quiet. It
contradicts the high level of magnetic activities required to
power energetic FRBs. Instead, they are interpreted as the orbital
motion of binary systems
\citep{zhang2020What,ioka2020binarya,li2021Periodic,sridhar2021Periodic,geng2021Repeating,kurban2022Periodic}
or the precession of neutron stars
\citep{yang2020Orbitinduced,tong2020Periodicity,levin2020Precessing,zanazzi2020Periodic,chen2021Reconciling},
or even the precession of the accretion discs
\citep{katz2022Precessiona}. A recent study indicates that the
long period is unlikely due to the free precession of a magnetar,
which should be damped in a few months \citep{Desvignes2024}.

Interestingly, a sub-millisecond quasi-periodic structure was
recently reported in five pulses spanning $\sim$ 2.14 ms for FRB
20201020A \citep{PastorMarazuela2023}. However, the
possibility that these pulses are simply sub-pulses of a longer
burst could not be ruled out. Alternatively, they might be
signatures of magnetar oscillations, whereby all pulses are
substructures of a single burst event comprising crustal
oscillations \citep{Wadiasingh2020}, or arise from multiple
emission regions regularly spaced within the magnetosphere.

Magnetars are highly magnetized neutron stars that occupy the
upper-right region of the $P$--$\dot P$ diagram. Their spin periods
are generally longer than those of ordinary radio pulsars, mainly
ranging from 1 to 12 seconds \citep{olausen2014McGill}. Recently
discovered long-period radio transients, with periodicities ranging
from hundreds to thousands of seconds, have further broadened the
phenomenology of radio emission from compact objects \citep{Rea2026}.
Some of these sources have been interpreted as possible ultra-long-period
magnetars, potentially produced through enhanced spin-down mechanisms
such as giant-flare kicks, charged-particle winds, or fallback
accretion \citep{Beniamini2023}. Alternative interpretations, including
magnetized white dwarfs and compact objects in binary systems, have also
been proposed \citep{Qu2025}. In short, the nature of these ultra-long-period
radio transients remains under debate. Given their substantial differences
from the FRB population in both luminosity and duration, caution is
required when interpreting them as FRB progenitors.

Notably, for the range from milliseconds to thousands of seconds,
extensive and exhaustive searches for periodicity in repeating
FRBs, particularly in active repeaters with thousands of detected
bursts, have all yielded null results
\citep{li2021bimodal,xu2022fast,niu2022FAST,Du2024}. The apparent
absence of periodicity connected to neutron star rotation has thus
posed a puzzle for the magnetar origin models. Some mechanisms
could alleviate this difficulty. For instance, the bursts may not
be produced at one or two fixed locations such as the dipolar cap
regions, but rather from multiple emission sites across the
magnetosphere (e.g., in a multi-polar field geometry or from
multiple magnetic reconnection regions). Furthermore, even in the
framework of dipolar models, if the inclination angle between the
rotational axis and magnetic axis is very small, then a very
low-amplitude modulation (or no periodicity) would be observed
when our line of sight is close to the magnetic axis. Anyway, the
discovery of short timescale periodicity would be the long-sought
proof to unambiguously link a repeating FRB to a rotating central
engine.

FRB 20201124A is an extremely active repeater, with nearly 3000
bursts being detected
\citep{hilmarsson2021Polarizationa,marthi2021Burst,xu2022fast,zhou2022FAST}.
It is located in a barred spiral galaxy at a redshift of $z$ =
0.098
\citep{fong2021Chronicling,piro2021fast,Ravi2022,xu2022fast}.
Although previous studies have reported some millisecond
quasi-periodicities at $\sim$ 3$\sigma$ level in the sub-pulses of
some multi-component bursts from this repeater
\citep{niu2022FAST}, all attempts to search for a rotation
periodicity in the source have yielded null results. In this
study, we conduct systematic period searches on the bursts from
FRB 20201124A, paying special attention to short-timescale
periodicity possibly associated with the rotation of a magnetar.

\section*{Results}
\label{sec:results}
\subsection*{Second-scale periodicity on MJDs 59310 and 59347}

Our full sample consists of 2812 bursts spanning 49 observing days and forming 51 single-day datasets (45 from FAST \#1, 4 from FAST \#2, 1 from uGMRT, and 1 from Effelsberg). Here, a single-day dataset is defined as the set of bursts detected by a single telescope on a single MJD date. The
largest number of events were observed on MJD 59485, with 542
bursts detected by FAST, while the smallest number of events were
observed on MJD 59344, with 11 bursts detected by FAST. We
performed a blind period search on each day separately. For an
initial screening, we adopt a reference detection threshold of
$3\sigma$ level based on the $\chi^2$ distribution. No significant
periodicity was found on 47 of the 49 observing days. The
periodograms for all days are shown in Supplementary
Figure 1. Furthermore, we conducted combined searches on
data grouped into 5-day blocks and still found no significant
periodicity.

However, our blind search revealed a clear periodicity on two
days, with candidate periods of $1.70603$ s on MJD 59310 and
$1.70797$ s on MJD 59347. To get the best-fit period and estimate
the uncertainty, we further performed a parametric bootstrap
analysis (see Methods), which yielded a refined period of $P =
1.706024(13)$ s on MJD 59310 and $P = 1.707968(9)$ s on MJD 59347
(see Figure~\ref{Fig7}).

On MJD 59310, 28 bursts were detected during a single observing session lasting for 7200 s, and the time span between the first
and last burst was 5527 s. When folded at the refined period, 86\%
(24/28) of the bursts cluster within a $\sim$ 35\% phase window.
On MJD 59347, 25 bursts were detected across four observing sessions, each lasting for 1800 s, and the time span between the
first and last burst was 12111 s. When folded at the refined
period, all bursts (25/25) cluster within a $\sim$ 45\% phase
window. The periodograms and the folded phase histograms for these
two days are shown in Figure~\ref{Fig2}. On MJD 59310, an
additional peak ($\sim$ 3.4 s) is also seen, which is a harmonic
of the $\sim$ 1.7 s period.

The significance levels of these two candidate periods are then
evaluated using extensive end-to-end Monte Carlo simulations that
reproduce the blind-search procedure and account for the trials factor. For the 1.70603 s period on MJD 59310, the
significance level is $3.9\sigma$, and for the 1.70797 s period on
MJD 59347, the significance level is $3.1\sigma$. We further
cross-checked the significance of these two periods by using the
$H$-test, which yields a significance level of $3.4\sigma$ and $5.2\sigma$ for
MJDs 59310 and 59347, respectively, thereby confirming the robustness of
the detected periodicity. Notably, the two periods are very close
to each other, which enhances the credibility of the detection and
suggests that they may reflect an intrinsic timescale of the
central engine. A global significance analysis, considering
both the periodic signal power and the consistency of the
two periods, yields a global significance level of $5.5\sigma$ (See
Methods for details of the global significance analysis).

The dynamic spectra of bursts are available in Ref. \citep{wang2023Atlas}, from which we can extract the pulse profile
of each burst on MJDs 59310 and 59347. This allows us to plot a
more intuitive folded phase diagram that illustrates the TOAs and pulse profiles of the bursts. The results are shown in
Figure~\ref{Fig3}, where the intensity of each burst is reflected by referring to its signal-to-noise ratio.

% \section*{Implications}
% \label{sec:implications}

% It is generally believed that FRBs should originate from neutron
% stars, or more specifically, magnetars. The 1.7 s periodicity
% of FRB 20201124A may correspond to the rotation of the compact
% central engine. In this section, we present some further
% discussions on this issue.

\subsection*{Magnetar origin}

FRB 20201124A is one of the most active repeaters observed to
date. Note that in this study, the periodicity is found on two
days that are 37 days apart ($P = 1.706024(13)$ s on MJD 59310 and $P
= 1.707968(9)$ s on MJD 59347). Even more astonishing is the fact
that these two periods are quite similar. The period also shows an
increasing trend. Assuming a linear evolution of
the period over the 37-day baseline, we obtain the period derivative as
$\dot{P}= 6.11(5) \times 10^{-10} \,\mathrm{s}
\,\mathrm{s}^{-1}$, which is interestingly quite typical for a
young magnetar.

Adopting the magnetic dipole radiation mechanism for
the spin down, we can further calculate the surface magnetic field
as
\begin{equation}\label{Eq1}
\begin{aligned}
B_{\mathrm{surf}} & = \left( \frac{3 I c^3 P \dot{P}}{8 \pi^2 R^6}
\right)^{\frac{1}{2}} \\
& \approx 1.03 \times 10^{15}\ \mathrm{G} \left( \frac{I}{10^{45}\
\mathrm{g}\ \mathrm{cm}^2} \right)^{1/2} \left( \frac{R}{10^6\
\mathrm{cm}} \right)^{-3},
\end{aligned}
\end{equation}
where $c$ is the speed of light. $I$ and $R$ are the moment of
inertia and radius of the neutron star, respectively. Here, we
adopt typical values for the neutron star's moment of inertia, $I
= 10^{45}\,\mathrm{g}\,\mathrm{cm}^2$, and radius, $R =
10^6\,\mathrm{cm}$. The derived magnetic field of
$B_{\mathrm{surf}} \approx 1.03 \times 10^{15} $ G further
supports that the central engine is a magnetar, whose spin-down
age is $\tau = P/(2\dot{P})\approx$ 44 yr. It is younger than all
previously known magnetars. To compare it with other pulsars in
the ATNF Pulsar Catalogue \citep{manchester2005Australia}, we have
plotted them on the $P$--$\dot{P}$ diagram, as shown in
Figure~\ref{Fig4}. We see that FRB 20201124A has a relatively high
surface magnetic field even among the magnetar population.

Note that some other observations also support the idea that the central engine of FRB 20201124A may be a magnetar. For example, orthogonal jumps in the polarization angle were recently reported in three bursts from FRB 20201124A \citep{niu2024Sudden}. Similar jumps are commonly observed in pulsars and are interpreted as transitions between two orthogonal polarization modes. Such behavior is broadly consistent with a rotating magnetosphere scenario, potentially involving propagation effects in a strongly magnetized magnetospheric plasma. Additionally, a persistent radio source is detected at the position of FRB 20201124A \citep{bruni2024nebular}, which could be a nebula associated with the young magnetar.

It is worth mentioning that quasi-periodic sub-pulse structures
are frequently observed in radio pulsars
\citep{Kramer2024}. In many well-studied radio-loud
magnetars, there is a simple linear correlation between the
quasi-period of sub-pulse structures and the rotation period, i.e.
 $P_{\mu}(\mathrm{ms}) = (0.94\pm0.04)\times P(\mathrm{s})
^{0.97\pm0.05}$ \citep{Kramer2024}. As a result, the
quasi-periodic sub-pulses in FRBs can also be used to infer the
underlying rotation period of the associated magnetar. Recently,
in a 150 ms burst from FRB 20201124A that contained 11 sub-pulses,
a 3.06 ms quasi-periodicity was identified at a confidence level
of 3.3$\sigma$ \citep{niu2022FAST}. Using the above correlation
obtained in Ref. \citep{Kramer2024}, we find that the rotation
period of FRB 20201124A would be estimated as
$3.38^{+0.41}_{-0.32}$ s, which happens to be nearly twice the
period we identified in this study. It thus raises an interesting
possibility that the intrinsic period of FRB 20201124A might be
$\sim$ 3.4 s rather than $\sim$ 1.7 s. When the bursts on MJDs
59310 and 59347 are folded by using the $\sim$ 3.4 s period, two
cluster windows separated by a phase of about 0.5 periods will be
observed in each folded profile.

We further quantified whether the apparent agreement between the 3.06 ms quasi-periodicity reported in Ref. \citep{niu2022FAST} and the possible 3.4 s harmonic of our candidate period could arise by chance. We constructed an empirical background distribution from the 54 quasi-periodicities reported in Ref. \citep{niu2022FAST} and calculated the probability for one quasi-periodicity drawn from this distribution would match the $\sim$ 1.7 s or $\sim$ 3.4 s period. The resulting chance probability is 0.06. After accounting for the 54 reported quasi-periodicities, the probability that at least one of them matches the $\sim$ 1.7 s or $\sim$ 3.4 s period becomes 0.96. Therefore, the apparent agreement is not statistically compelling, and we do not use it as independent support for our periodicity.

\subsection*{Randomness of bursts in the active window}

From Figure~\ref{Fig3}, we can also see that strong bursts and
weak bursts occur randomly in the active window, i.e. there is no
clear tendency for strong/weak bursts to cluster at any particular
phase. Other random behaviors of FRBs were also reported in Ref.
\citep{zhang2024arrival}, which is different from the radio
emission of pulsars. The randomness of bursts in the active window
indicates that FRBs may come from different emission zones in the
magnetosphere of the magnetar.

Some methods of periodicity analysis, such as the Fast Fourier
Transformation, Lomb-Scargle periodogram, and the autocorrelation
function method, heavily rely on the intensity information of each
burst, since some kinds of phase-dependent properties of the
bursts are assumed in the analysis. The randomness of FRBs
discovered in our study strongly indicates that these methods are
not appropriate in the periodicity analysis of FRBs. In fact, we
have also tried to use the Lomb-Scargle periodogram to search for
periodicity in the bursts on MJDs 59310 and 59347, adopting the
flux of each burst as the necessary input of intensity
information. No periodicity is found at around 1.7 s. We thus
stress that the phase-folding method, which does not depend on
intensity information, still seems to be the most appropriate
operation for periodicity analysis of repeating FRBs. Notably, a
recent study in Ref. \citep{Shen2026} successfully recovered our
reported $\sim$ 1.7 s period using an independent method that
combines phase folding with Markov Chain Monte Carlo parameter
estimation, supporting the periodicity on the two days.

\subsection*{Non-detection of clear periodicity on other days}

Significant periodicity is detected only on MJDs 59310 and 59347. Our blind period searches yielded null results on other days.
Figure~\ref{Fig5} compares the properties of the bursts on MJDs
59310 and 59347 with other bursts on the width-fluence plane. We
see that there is no systematic difference between these two
groups of events.

A plausible interpretation for the absence of
clear periodicity on most days is that the burst activity is not
always confined to a single phase window. In other words, bursts
are not tied to one special site in the magnetosphere. Instead,
they may originate from multiple emission sites. In this case, the
periodicity connected to the spin of the magnetar will be
effectively smeared when all these sites actively give birth to
bursts on most days.

To search for potential observational evidence for this
interpretation, we used a linear period-evolution model (see
Methods) to extrapolate the two detected periods to other days. We
then folded the bursts detected on each day using the extrapolated
period. For some days, although no significant periodicity is
detected in the blind search, folding the bursts using the
extrapolated period nevertheless reveals interesting structure in
phase. On some days, there are even multiple peaks in the profile,
as shown in Figure~\ref{Fig6}. In fact, each peak
could be indicative of an emission site so that multiple emission
sites are interestingly favored. The result thus further strongly
supports the authenticity of the periodicity on MJDs 59310 and
59347, and it is qualitatively consistent with the multiple
emission site interpretation as well.

We have conducted simulations to test the effects of multiple emission sites on period detection. In our
simulations, $N$ mock bursts are generated in a time span of 2 hr,
which are assumed to originate from $m$ different sites in the
magnetosphere of a rotating magnetar with a 1.7 s period. Here $N$
is taken as $N = 10, 20, 30, \ldots, 100$ and $m$ is taken as $m =
1, 2, 3, \ldots, 10$. It is found that when the source is in a
highly active phase (e.g., 100 bursts detected in 2 hr), periodicity becomes undetectable when bursts
originate from multiple emission sites with $m \geq 2$. On the
other hand, when the source is in a quiescent phase (e.g., only 20--30 bursts detected in 2 hr), periodicity is
detectable only if most bursts come from a single emission site
($m = 1$). In the case of FRB 20201124A, it is quite possible that
only one dominant emission site was active on MJDs 59310 and 59347, while
all other emission sites were quenched due to unknown reasons. As
a result, we could only detect the significant periodicity on these two days. On
the contrary, a number of emission sites may be actively producing
FRBs simultaneously on other days, leading to the smearing of the
periodicity. However, the mechanism of turning on and turning off
of the emission sites is still to be explored.

A recent theoretical work of Ref. \citep{Zhang2025} offers a
possible explanation for the switching on/off mechanism. It is
argued that the central engine of FRB 20201124A is a magnetar in a
binary system, with a Be-star companion. When the magnetar crosses
the wind disk of the Be star at two specific geometric positions
(on MJDs 59310 and 59347), the disk-magnetar interaction could
suppress the magnetar's lower-latitude, multi-polar emission. This
would leave only the primary polar region active, reasonably
explaining the temporary appearance of a strong, clean periodic
signal on the two days.

% \section*{Summary}

% \label{sec:summary}

% In this study, we have collected 2812 bursts of FRB 20201124A
% generated on 49 days. Periodicity analysis is performed by using
% the phase-folding method on the bursts of each day. A clear
% periodicity of $1.706024(13)$ s is identified for the 28 bursts on
% MJD 59310, and a periodicity of $1.707968(9)$ s is also found for
% the 25 bursts on MJD 59347. A global Monte Carlo analysis based on all single-day datasets yields a significance level of
% $5.5\sigma$. The increase of the period indicates a derivative of
% $6.11(5) \times 10^{-10} \ \mathrm{s} \ \mathrm{s}^{-1}$. A
% surface magnetic field strength of $1.03 \times 10^{15}\
% \mathrm{G}$ is estimated under the dipolar magnetic field
% assumption, strongly hinting a young magnetar as the central
% engine. It is argued that FRBs could originate from multiple sites
% in the magnetosphere of the magnetar, which effectively smears the
% periodicity of bursts on other days. On MJDs 59310 and 59347, only
% one dominant emission sites should be active while other sites were
% quenched due to some unknown reasons, leaving a clear periodicity
% signal in the bursts. Our study provides the first clear evidence
% for the spinning of the central engine and hints that FRB
% 20201124A originates from a young magnetar.

\section*{Methods}
\label{sec:methods}

\subsection*{Data}

A large number of bursts were detected from FRB 20201124A during
its two active episodes in 2021. In the first episode ranging from
April to June 2021, FAST detected 1863 bursts in a total
monitoring time of 88 hours over 45 days, with a peak burst rate
of 46 hr$^{-1}$ \citep{xu2022fast}. A few months later, during
another extremely active episode in September, FAST observed 881
bursts in a total duty time of 4 hours over 4 days, with a peak
burst rate reaching an astonishing 542 hr$^{-1}$
\citep{zhou2022FAST}. In this study, we mainly utilize the bursts
detected by FAST for periodicity analysis. To expand the sample
size as far as possible, some bursts detected by other instruments
are also included.

During the first active episode of FRB 20201124A (from April to
June, 2021), the bursts were observed by FAST with its L-band (1.0--1.5 GHz) 19-beam receivers. The data were recorded with a
temporal resolution of 49.152 $\mu$s or 196.608 $\mu$s and the
frequency resolution is 122.07 kHz. The TOAs of each burst were
measured with respect to the centroid of the best-matched boxcar
filter and were corrected to the barycentric time at 1.5 GHz
\citep{xu2022fast}. The pulse profiles of the bursts are extracted
from the full dynamic spectra \citep{wang2023Atlas}, using the \texttt{PSRCHIVE} code \citep{hotan2004PSRCHIVE}. The specific dedispersion
strategy and signal-to-noise ratio threshold used for burst
searches are detailed in Refs. \citep{xu2022fast,wang2023Atlas}.

The bursts during the second active episode of 2021 (September 25--28) were also recorded by FAST's L-band receivers. The data
have a temporal resolution of 49.152 $\mu$s and a frequency
resolution of 122.07 kHz. The TOAs of the bursts are also
corrected to the barycentric times at 1.5 GHz
\citep{zhou2022FAST}. The data are directly available in Ref. \citep{zhang2022FASTa}.

\begin{figure*}
\centering
\includegraphics[width=1\textwidth]{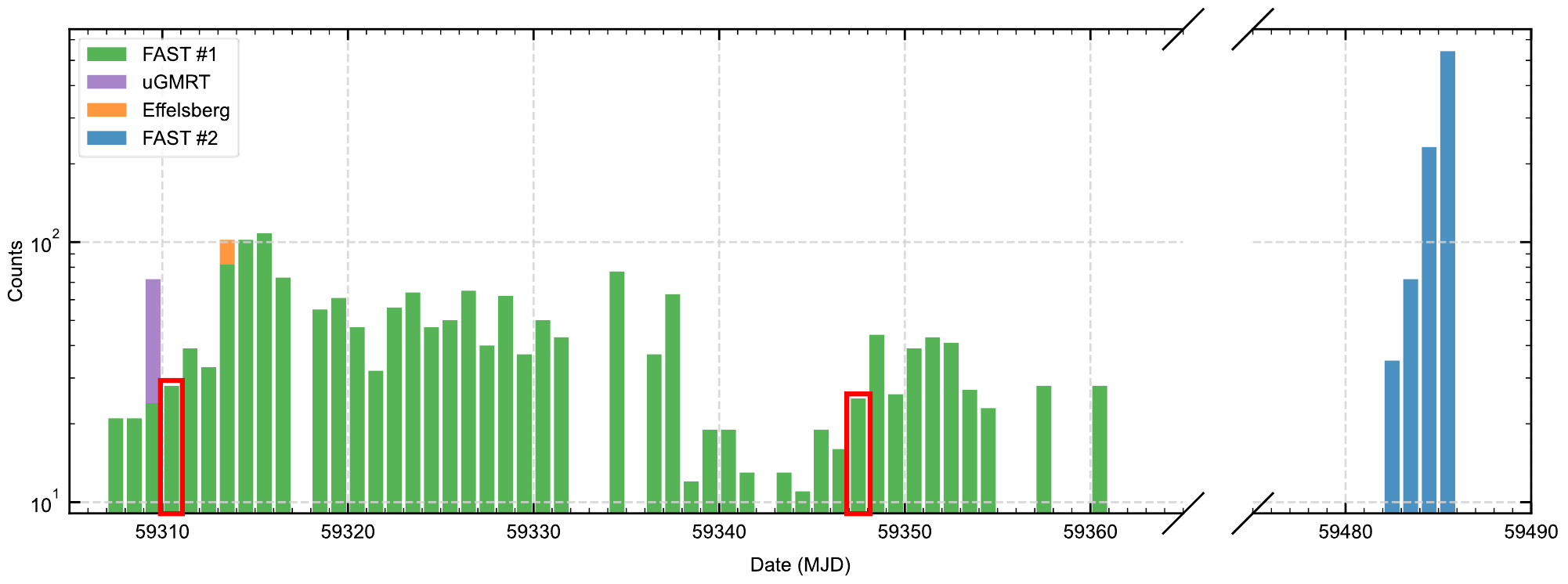}
\caption{{\bf Number of bursts detected on each day for FRB 20201124A.}
The X-axis represents the observing date (MJD), and the Y-axis
represents the number of bursts detected on that day. Bursts
detected by FAST during the first active episode are shown in
green, while the bursts of the second active episode are shown in
blue. The bursts detected by uGMRT (purple) and Effelsberg
(orange) are also included. The two days with periodicity detected
(MJDs 59310 and 59347) are marked by red boxes.} \label{Fig1}
\end{figure*}

\begin{figure*}
\centering
\includegraphics[width=1\textwidth]{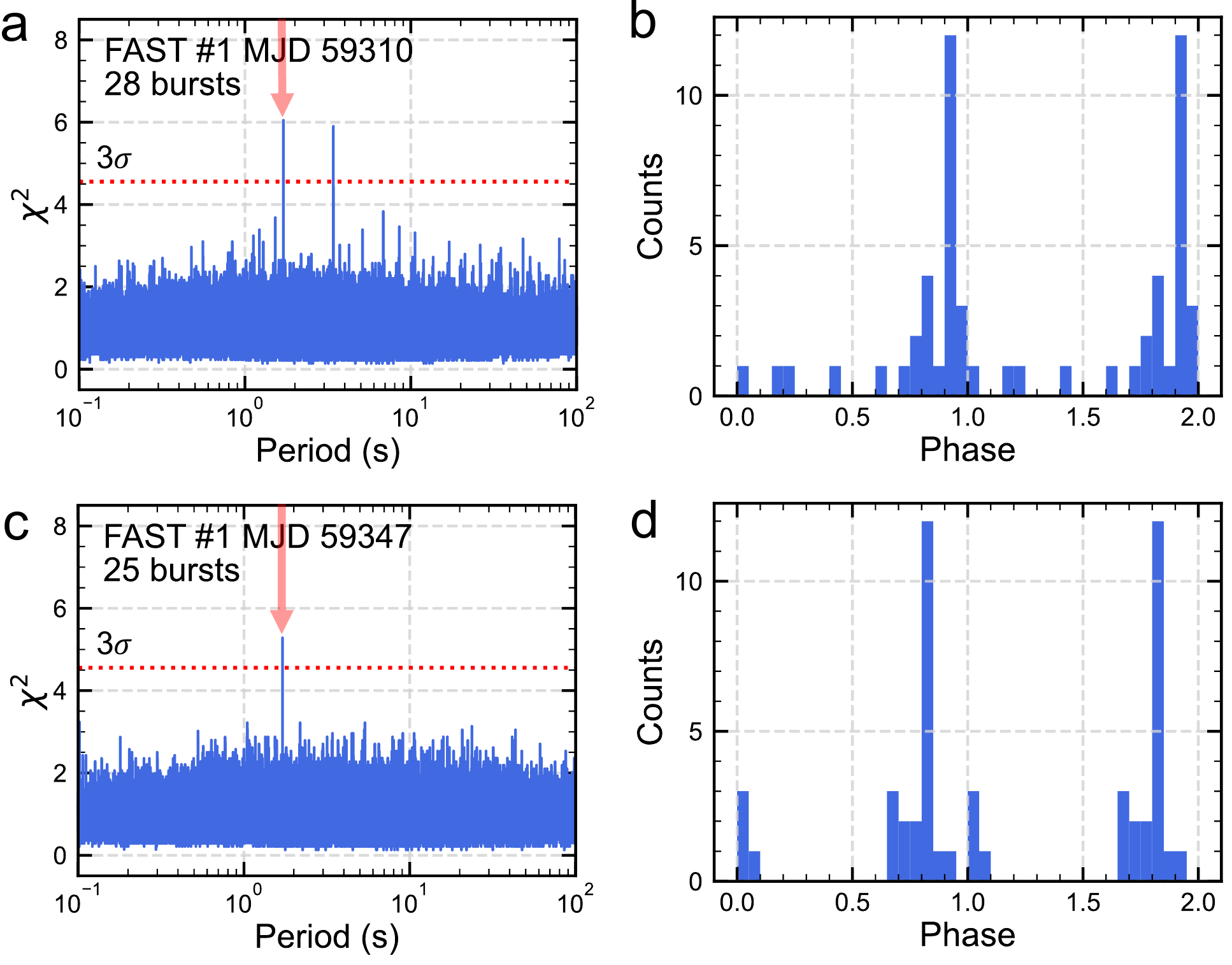}
\caption{{\bf The periodograms and the folded phase histograms for the
bursts on MJD 59310 (panels a and b) and MJD 59347 (panels c and d).}
Panels a and c show the $\chi^2$ periodograms, with the
horizontal dotted line indicating a significance level of
$3\sigma$ derived from the theoretical $\chi^2$ distribution. The
red arrows in the periodograms indicate the periods identified in
the blind search: 1.70603 s on MJD 59310 and 1.70797 s on MJD
59347. On MJD 59310, an additional peak ($\sim$ 3.4 s) is also
seen, which is a harmonic of the $\sim$ 1.7 s period. Panels b and d show the phase histograms for each day, folded by the
refined periods of 1.706024 s and 1.707968 s, respectively. For
clarity of visualization, two complete cycles are plotted. The
mean TOA of bursts observed on each day is taken as the time of
the zero phase point.}

\label{Fig2}
\end{figure*}

Other data utilized in our analysis include: 46 bursts
detected by uGMRT (0.55--0.75 GHz) on April 5, 2021
\citep{marthi2021Burst}, and 20 bursts detected by Effelsberg
(1.21--1.51 GHz) on April 9, 2021
\citep{hilmarsson2021Polarizationa}. The log of all the bursts in
our sample is illustrated in Figure~\ref{Fig1}. More
details of all observing sessions, including the specific start
and end times for each continuous observing session, are provided in
Supplementary Table~\ref{STab1}.

We have also conducted followup monitoring of FRB 20201124A with FAST.
Six one-hour observations were performed on August 8, September 8,
October 8, November 8, December 1, and December 20 2025. Two additional
30-minute observations were further carried out on January 10 and February 1
in 2026. We searched for bursts in each observation using the \texttt{PRESTO}
code \citep{Ransom2001}. Unfortunately, no new bursts were detected during the
total 7 hours of observations spanning approximately six months.
It seems that the source has entered a quiescent phase.

\subsection*{Phase folding method for periodicity analysis}
Various methods could be used for periodicity analysis, many of
which require the intensity information as necessary input. The
phase-folding method is a traditional yet effective and practical
method for analyzing periodicity. It only requires the TOAs of the
events. Given the incomplete and nonuniform nature of FRB data,
this method is more direct and reliable for our purpose. In our
study, we employ the phase-folding method to search for potential
periodicity. Bursts on each day are analyzed separately.

To begin the analysis, we first assume a trial period ($p$) for
the bursts and fold their TOAs by $p$. The normalized phase of a burst arriving at $t_i$ can then be calculated as
\begin{equation}\label{Eq2}
\phi_i = \frac{t_i\,\mathrm{mod}\,p}{p} ,
\end{equation}
where mod means a remainder operation. If $p$ were the intrinsic
period of the time series, a significant number of bursts will
cluster at specific phases. Otherwise, the bursts will be randomly
distributed across the whole phase range. By varying $p$ in a
particular time interval, we will be able to identify potential
periodicity in the target period range.

We use the classic Pearson's $\chi^2$ test to assess the
concentration of the bursts in the phase space. For a specific
trial period $p$, when the phases of all the bursts have been
determined, we group them into $n$ bins in the phase space. The
reduced $\chi^2$ value is then calculated as
\begin{equation}\label{Eq3}
\chi ^2=\sum_{j=1}^{n}\frac{(O_j-E_j)^2}{(n-1)E_j},
\end{equation}
where $O_j$ is the observed event count in the $j$-th bin, and
$E_j$ is the expected mean count in each bin for a uniform phase
distribution.

Usually, a large $\chi^2$ value indicates that $p$
could potentially be the period of the burst activity, while a
small $\chi^2$ means that $p$ is unlikely the intrinsic period.
Varying $p$ in all the possible period ranges, we would be able to
find the true period if it really exists. In our analysis, we set the number of bins as $n=20$.

Note that most single-day datasets could not satisfy an expected count
of $>5$ bursts per bin when performing the $\chi^2$ test. With
such a small sample size, the distribution might deviate from the
theoretical $\chi^2$ distribution. A common approach to alleviate
this problem is to merge the bins \citep{Wall2012}. However, a bin number less than
5 would drastically reduce the resolution of the search and can
easily miss structures in the diagram. To mitigate the small
sample size issue, we applied Williams' correction to the $\chi^2$
statistic \citep{Williams1976,McDonald2014}, which involves
dividing the original $\chi^2$ value by a correction factor $q$,
\begin{equation}\label{Eq4}
\chi ^2_{\mathrm{correct}}=\frac{\chi ^2}{q}=\frac{\chi^2}{1+(n+1)/6N},
\end{equation}
where $N$ is the sample size. It means that for a fixed number of
bins, the correction factor is larger for smaller sample sizes,
causing the corrected $\chi^2$ distribution to more closely
approximate the theoretical distribution. It effectively
alleviates the small sample size problem.

\subsection*{Period search strategy}

In our study, we aim to search for short-timescale periodicity. We
therefore analyze each single-day dataset separately. The primary
motivation is that any plausible period derivative would be
negligible over the duration of a single-day observation
(typically no more than 3--4 hours), so the periodicity search
can be performed without modeling period evolution. As a secondary
benefit, single-day analyses mitigate potential day-to-day
systematic timing offsets introduced in the barycentric
correction. Finally, restricting the time span to single-day observations also reduces the computational cost of the blind period
search.

We use the phase folding method to analyze the periodicity of the
bursts detected on each day separately. The target period range is
0.1--100 s. A period step of $10^{-5}$ s is adopted.

We have tested our search strategy by means of numerical
simulations. For this purpose, a number of mock TOA sequences were generated. These sequences contain different numbers of FRBs over different observing time spans, with different pre-assumed intrinsic periodicities. Observational truncation
effects during the data acquisition are also considered, and
noises (i.e., random offsets) are injected into the TOAs.
The aforementioned periodicity analyzing strategy was applied to
the mock TOA sequences. In all the cases, the pre-assumed
periodicity can be correctly recovered, which clearly proves the
effectiveness of our method.

It is worth noting that the waiting time of repeating FRBs usually
exhibits a bimodal log-normal distribution. Typically, the peak
waiting time of the shorter component is of the order of tens of
milliseconds. Such closely spaced bursts can interfere with our
searches for long periods. Therefore, in our analysis we apply a
de-clustering threshold of 0.5 s when searching for periods longer
than 1 s. Specifically, bursts separated by less than 0.5 s are
grouped into the same cluster, and the burst with the highest peak
flux within each cluster is selected as the cluster
representative. All the single-day period search results are shown
in Supplementary Figure~\ref{SFig1}.

\subsection*{Refining the period with bootstrap analysis}

Our blind search revealed a clear periodicity on two days, i.e.,
$P_{\rm candidate}=1.70603$ s on MJD 59310 and $P_{\rm
candidate}=1.70797$ s on MJD 59347. To further constrain the exact
period value and the corresponding error range, we have performed
a more in-depth parametric bootstrap analysis on the bursts of
these two days.

First, for the observed TOA series on these two days, we performed a
local period search within a narrow range ($\pm 10^{-4}$ s) around
$P_{\rm candidate}$ with a step size of $10^{-8}$ s and obtained
an initial estimation for the period ($\hat{P}_0$). On MJD 59310,
we got $\hat{P}_0 = 1.70602821$ s, and on MJD 59347, we got
$\hat{P}_0 = 1.70797064$ s.

The bootstrap procedure then started by folding the observed TOAs
with $\hat{P}_0$. For each TOA $t_i$, we decomposed it into an
integer number of cycles $r_i$ and a phase $\phi_i$
($\phi_i\in[0,1)$),
\begin{equation}\label{Eq5}
t_i = (r_i+\phi_i)
\cdot\hat{P}_0.
\end{equation}
The set of phases $\{\phi_i\}$ characterizes the phase
distribution of the observed bursts.

We divided the whole phase range of 0--1 into $n$ = 20 bins. For
the set of phases $\{\phi_i\}$, we counted the number of events in
each bin and denoted the resulting count distribution as ${\bf
c}=(c_1,\dots,c_{20})$. Consequently, the likelihood of observing
the counts ${\bf c}$ is modeled by a multinomial distribution,
${\bf c}\,|\,{\bf p}\sim \mathrm{Multinomial}(N,{\bf p})$, where
$N$ is the total number of events and ${\bf p}=(p_1,\dots,p_{20})$
denotes the probabilities of an event falling into each bin. We
adopted a symmetric Dirichlet prior for ${\bf p}$,
\begin{equation}\label{Eq6}
{\bf p}\sim \mathrm{Dirichlet}(\alpha_0,\dots,\alpha_0),
\end{equation}
where we set $\alpha_0=0.5$ to avoid zero-probability bins. Since the Dirichlet distribution is conjugate to the multinomial distribution \citep{gelman1995bayesian}, the posterior distribution of ${\bf p}$ is also a Dirichlet distribution, with parameters updated by adding the observed counts,
\begin{equation}\label{Eq7}
{\bf p}\,|\,{\bf c} \sim \mathrm{Dirichlet}(\alpha_0+c_1,\dots,\alpha_0+c_{20}).
\end{equation}

In the $k$-th bootstrap realization, we draw a phase profile ${\bf
p}^{(k)}$ from the posterior in Eq.~(\ref{Eq7}). We then generated
a synthetic phase $\phi_i^{(k)}$ by choosing a phase bin according
to ${\bf p}^{(k)}$ and sampling uniformly within that bin.

Finally, we generated a bootstrap TOA series by keeping the
observed integer number of cycles fixed and replacing only the
phases,
\begin{equation}\label{Eq8}
t_i^{(k)} = \left(r_i+\phi_i^{(k)}\right)\cdot\hat{P}_0.
\end{equation}
Additionally, we added a Gaussian timing jitter drawn from
$\mathcal{N}(0,\sigma_t^2)$ to each $t_i^{(k)}$, with
$\sigma_t=1\times10^{-4}$ s.

For the $k$-th bootstrap sample, we applied the same de-clustering
procedure used in our actual search pipeline and then obtained the
period $\hat{P}^{(k)}$ by performing the same local search within
a narrow range ($\pm 10^{-4}$ s) around $P_{\rm candidate}$ with a
step size of $10^{-8}$ s. For each of MJDs 59310 and 59347, we
generated $10^5$ bootstrap samples. For the resulting bootstrap
distribution $\{\hat{P}\}$, we defined the $\pm1\sigma$
uncertainties using the equal-tailed central 68\% interval,
\begin{equation}\label{Eq9}
\sigma_{-} = \hat{P}_{50\%}-\hat{P}_{16\%},\qquad
\sigma_{+} = \hat{P}_{84\%}-\hat{P}_{50\%}.
\end{equation}
Here $\hat{P}_{16\%}$ and $\hat{P}_{84\%}$ are the 16th and 84th
percentiles of the distribution, respectively, and
$\hat{P}_{50\%}$ is the median of the distribution. We adopted
$\hat{P}_{50\%}$ as the final refined period and derived a
symmetric uncertainty of $\sigma=(\sigma_{+}+\sigma_{-})/2$.

Using the above parametric bootstrap method, the final period is
refined as 1.706024(13) s on MJD 59310,  and 1.707968(9) s on MJD
59347, respectively. Note that these values only differ slightly
from the corresponding $\hat{P}_0$ value. The distributions of
$\{\hat{P}\}$ for MJDs 59310 and 59347 are displayed in Figure~\ref{Fig7}, with the $\pm1\sigma$ bounds indicated.

\subsection*{Quantitative analysis of periodograms}
We collected all $\chi^2$ values from the periodograms shown in Supplementary Figure~\ref{SFig1} and constructed their distribution, which is shown in Supplementary Figure~\ref{SFig2}. We overplotted the theoretical $\chi^2$ curve on the distribution. The comparison shows that the bulk of the $\chi^2$ distribution is broadly consistent with the theoretical $\chi^2$ curve. Importantly, the peak $\chi^2$ values on MJDs 59310 and 59347 lie in the extreme sparse tail of the distribution. Rather than representing a natural extension of the high-value tail of the noise distribution, they stand out as distinct outliers.

We further examined the chance probability of finding such close peak pairs directly in the real periodograms. For each periodogram, we identified local peaks with $\chi^2 > 2.5$. For any two distinct single-day datasets $a$ and $b$, the total number of peak pairs is given by
\begin{equation}\label{Eq10}
M_{\rm pair}=\sum_{a<b} n_a n_b,
\end{equation}
where $n_a$ and $n_b$ denote the number of peaks in the periodograms for single-day datasets $a$ and $b$, respectively.

For a given detection threshold $X$ and maximum allowed period separation $\Delta P_{\max}$, we counted the number of peak pairs satisfying
\begin{equation}\label{Eq11}
\chi^2_{a,u}\geq X,\qquad
\chi^2_{b,v}\geq X,\qquad
|P_{a,u}-P_{b,v}|\leq \Delta P_{\max},
\end{equation}
where $u,v$ denote local peaks within the corresponding periodograms. We denote this count as $C_{\rm pair}(X,\Delta P_{\max})$.
Then we obtain the chance probability $f_{\rm pair}$
\begin{equation}\label{Eq12}
f_{\rm pair}(X,\Delta P_{\max})
=
\frac{C_{\rm pair}(X,\Delta P_{\max})}{M_{\rm pair}}.
\end{equation}
For a grid of $X$ and $\Delta P_{\max}$, we obtain a two-dimensional heat map, as shown in Supplementary Figure~\ref{SFig3}. At the point corresponding to the lower of the two candidate peak powers and the observed period separation between MJDs 59310 and 59347, we find $C_{\rm pair}=1$, i.e., only the peak pair identified in the periodograms of MJDs 59310 and 59347 satisfies both the detection-threshold and period-separation criteria. The corresponding chance probability is $f_{\rm pair}=1.79\times10^{-9}$, which is directly measured from the real periodograms. The periodicity has a formal global significance of $5.5 \sigma$, as estimated using the end-to-end Monte Carlo procedure described below.

\subsection*{Statistical significance of the periodicity}

We estimated the significances of the observed periods on MJDs
59310 and 59347 through end-to-end Monte Carlo (MC) simulations based on all 51 single-day datasets (45 from FAST \#1, 4 from FAST \#2, 1 from uGMRT, and 1 from Effelsberg) included in our analysis. For each single-day dataset, we generated null-hypothesis TOA sequences
using the actual observing time span and the observed number of bursts (see Supplementary Table~\ref{STab1}).
We then applied the
same period-search pipeline as used for the real data, covering the
period range of 0.1--100 s with a step size of $10^{-5}$ s. As in the
real-data search, the same de-clustering procedure was applied when
searching for periods longer than 1 s.
To account for trials factor in the whole 0.1--100 s period
range, we recorded the maximum $\chi^2$
value obtained over the entire period range. For each single-day dataset, we simulated $10^5$ samples to build the null distribution
of $\chi^2_{\max}$. Since different single-day datasets have different observing time spans and different numbers of bursts, their null hypotheses are different, and therefore their raw $\chi^2_{\max}$ distributions are not directly comparable. To place all single-day datasets on a common significance scale, we converted any candidate value $x$ into a tail probability within each single-day dataset,
\begin{equation}\label{Eq13}
P(\chi^2_{\max} \geq x)
=
\frac{
N(\chi^2_{\max} \geq x)
}{
N_{\rm local}
},
\end{equation}
where $N(\chi^2_{\max}\geq x)$ is the number of simulations in which $\chi^2_{\max}$ exceeds $x$, and
$N_{\rm local}=10^5$ is the total number of simulations for each single-day dataset.

We then defined a significance score $\mathcal{Q}$ as
\begin{equation}\label{Eq14}
\mathcal{Q}
=
-\log_{10}
P(\chi^2_{\max} \geq x) .
\end{equation}
Because $\mathcal{Q}$ is normalized by the null distribution of each single-day dataset, it provides a common significance scale. The simulated $\mathcal{Q}$ values from all single-day datasets were then pooled to construct a combined null distribution, against which the significances of periods on MJDs 59310 and 59347 were evaluated. On MJD 59310, the significance level was found to be $3.9\sigma$. On MJD 59347, the significance level was found to be $3.1\sigma$. Figure~\ref{Fig8} shows the distribution of the $\mathcal{Q}$ statistic, with the values for MJDs 59310 and 59347 indicated.

Additionally, we employed the $H$-test to cross-examine the
identified periodicity. We searched the period range of 0.1--100
s with a step of $10^{-5}$ s, and set the maximum number of
harmonics to 20 to maintain sensitivity to complex phase profiles.
The $H$-test also independently confirmed the periodicity on the
two days, consistent with the $\chi^2$ test results. Specifically,
it gives a period of 1.70604 s on MJD 59310, and 1.70797 s on MJD
59347. Using the empirical relation of $P(>H)=\exp(-0.4H)$
\citep{deJager2010} and applying a trials factor correction for
the $\sim 10^{7}$ period trials, we obtained a Gaussian
significance level of $3.4\sigma$ and $5.2\sigma$ on MJDs 59310
and 59347, respectively.

The credibility of the periodicity is further enhanced by the fact
that the two periods are very close to each other and
imply a period derivative typical of a young magnetar. We
therefore performed a global analysis based on all single-day datasets to evaluate the probability that two such significant and closely spaced periods could arise by chance.

To this end, we used the local null simulations of all 51 single-day datasets to construct $N_{\rm global}$ global MC simulations. In each simulation, we randomly selected one local null simulation from each of the 51 single-day datasets and then assembled them into one mock burst sample. In this way, each global simulation yields a mock burst sample containing 51 mock single-day datasets, matching the structure of the real analyzed sample.

Within each global simulation, for any pair of mock single-day datasets $a$ and $b$ ($1\leq a<b\leq 51$), we first define the pair-significance score $\mathcal{R}$ as
\begin{equation}\label{Eq15}
\mathcal{R}
=
\min\left[
\frac{\max(\mathcal{Q}_a,\mathcal{Q}_b)}{\mathcal{Q}_{\rm obs,strong}},
\frac{\min(\mathcal{Q}_a,\mathcal{Q}_b)}{\mathcal{Q}_{\rm obs,weak}}
\right],
\end{equation}
where $\mathcal{Q}_{\rm obs,strong}$ and $\mathcal{Q}_{\rm obs,weak}$ are the larger and smaller $\mathcal{Q}$ values of the two observed periods in MJDs 59310 and 59347, and $\mathcal{Q}_a$ and $\mathcal{Q}_b$ are the $\mathcal{Q}$ values of the candidate
periods in mock single-day datasets $a$ and $b$, respectively. Thus, $\mathcal{R}\geq1$ means that the two candidate periods in a given pair of mock single-day datasets are at least as significant as the periods in the observed MJD 59310--59347 pair.
We also define a period-closeness score $\mathcal{S}$ as
\begin{equation}\label{Eq16}
\mathcal{S}
=
\frac{\Delta P_{\rm obs}}{|P_a-P_b|},
\end{equation}
where $\Delta P_{\rm obs}$ is the separation of the two observed periods in MJDs 59310 and 59347, and $P_a$ and $P_b$ are the candidate periods of mock single-day datasets $a$ and
$b$. Thus, $\mathcal{S}\geq1$ means that the two candidate periods in a given pair of mock single-day datasets are at least as close as the periods in the observed MJD 59310--59347 pair.

We then combine these two scores by defining a statistic $\mathcal{T}$ as
\begin{equation}\label{Eq17}
\mathcal{T}=\min(\mathcal{R},\mathcal{S})
=
\min\left[
\frac{\max(\mathcal{Q}_a,\mathcal{Q}_b)}{\mathcal{Q}_{\rm obs,strong}},
\frac{\min(\mathcal{Q}_a,\mathcal{Q}_b)}{\mathcal{Q}_{\rm obs,weak}},
\frac{\Delta P_{\rm obs}}{|P_a-P_b|}
\right].
\end{equation}
For the
observed MJD 59310--59347 pair, $\mathcal{T}=1$ by definition. This statistic is conservative in the sense that a pair of mock single-day datasets reaches $\mathcal{T}\geq1$ only if its two candidate periods are both at least as significant and at least as close as the periods of observed MJD 59310--59347 pair.

For each global MC simulation, we computed $\mathcal{T}$ for all possible pairs of mock single-day datasets and recorded the maximum value $\mathcal{T}_{\rm max}$. The global false-alarm probability was estimated as
\begin{equation}\label{Eq18}
{\rm FAP}_{\rm global}
=
\frac{
N(\mathcal{T}_{\rm max}\geq1)
}{
N_{\rm global}
}.
\end{equation}

In our calculations, among $10^9$ global MC simulations, only 20 have
$\mathcal{T}_{\rm max}\geq1$. We therefore obtain ${\rm FAP}_{\rm global}=2\times10^{-8}$, corresponding to a Gaussian significance level of $5.5\sigma$. This global false-alarm probability accounts for the trials over all possible pairs of all single-day datasets. Figure~\ref{Fig9} shows the distribution of  $\mathcal{T}_{\rm max}$.

In addition, we performed a two-dimensional parameter scan to investigate how the false-alarm probability depends on the adopted pair-significance threshold and period-closeness threshold. For this purpose, we used the same global MC procedure described above. We considered a grid of pair-significance thresholds $\mathcal{R}_{\rm th}$ and period-closeness thresholds $\mathcal{S}_{\rm th}$. For each grid point, we counted how many global MC simulations in which at least one pair of mock single-day datasets satisfied
\begin{equation}\label{Eq19}
\mathcal{R} \geq \mathcal{R}_{\rm th},
\qquad
\mathcal{S} \geq \mathcal{S}_{\rm th}.
\end{equation}
We denote this number by $N(\mathcal{R}_{\rm th},\mathcal{S}_{\rm th})$. The corresponding false-alarm probability is then
\begin{equation}\label{Eq20}
{\rm FAP}(\mathcal{R}_{\rm th},\mathcal{S}_{\rm th})
=
\frac{
N(\mathcal{R}_{\rm th},\mathcal{S}_{\rm th})
}{
N_{\rm global}
}.
\end{equation}

The resulting probability map is shown in Figure~\ref{Fig10}. As expected, the probability decreases for
larger $\mathcal{R}_{\rm th}$ and $\mathcal{S}_{\rm th}$. The observed MJD 59310--59347 pair corresponds to $\mathcal{R}_{\rm th}=1$ and $\mathcal{S}_{\rm th}=1$, marked by the star symbol. The observed point lies in the low-probability region of the
map, and the neighbouring grid points also correspond to low
probabilities. This indicates that the inferred global significance
is not driven by a finely tuned choice of the pair-significance or
period-closeness threshold, and is therefore robust.

% #################################
% #################################

\subsection*{Phase folding analysis based on a linear period-evolution model}
For days on which the blind period search does not yield a significant periodicity, we further examine whether the bursts nevertheless exhibit a structured phase distribution when folded under the period-evolution model inferred from the two detected periods on MJDs 59310 and 59347.

We assume that the period evolves linearly in time,
\begin{equation}\label{Eq21}
P(t) = P_{\rm ref} + \dot{P}\cdot(t-t_{\rm ref}),
\end{equation}
where $P_{\rm ref}$ is the period at a chosen reference time $t_{\rm ref}$, and $\dot{P}$ is assumed to remain constant over the time span considered. In this framework, the accumulated rotational phase, measured in cycles, between $t_{\rm ref}$ and a burst TOA $t_i$ is
\begin{equation}\label{Eq22}
\phi_i = \int_{t_{\rm ref}}^{t_i} \frac{dt}{P(t)}
= \frac{1}{\dot{P}}
\ln\!\left[1 + \frac{\dot{P}\cdot(t_i-t_{\rm ref})}{P_{\rm ref}}\right].
\end{equation}

This linear period-evolution model is anchored by the two detected periods on MJDs 59310 and 59347, whose observed central values are $P_0 = 1.706024(13)$ s and $P_1 = 1.707968(9)$ s, respectively. Let $t_0$ and $t_1$ denote the mean burst TOAs on these two days. To propagate the uncertainties of the detected periods, we adopt a Monte Carlo sampling approach. In the $l$-th Monte Carlo realization, one value $P_0^{(l)}$ and one value $P_1^{(l)}$ are drawn from normal distributions centered on $P_0$ and $P_1$, with standard deviations set by the $1\sigma$ uncertainties. The corresponding sampled period derivative is then
\begin{equation}\label{Eq23}
\dot{P}^{(l)} = \frac{P_1^{(l)} - P_0^{(l)}}{t_1 - t_0},
\end{equation}
and the corresponding sampled period at the reference time $t_{\rm ref}$ of the target day is
\begin{equation}\label{Eq24}
P_{\rm ref}^{(l)} = P_0^{(l)} + \dot{P}^{(l)}\cdot(t_{\rm ref} - t_0).
\end{equation}
Here, $t_{\rm ref}$ is the mean burst TOA on that day.

Each Monte Carlo realization then yields $P_{\rm ref}^{(l)}$ and $\dot{P}^{(l)}$, and hence a corresponding set of folded phases $\{\phi_i^{(l)}\}$ for all bursts observed on that day. In practice, for each day, we generate $10^4$ Monte Carlo realizations, thereby obtaining an ensemble of phase distributions. To summarize and visualize this ensemble, we construct a phase histogram and a kernel-density estimate in each Monte Carlo realization, and then characterize the ensemble by its mean and $\pm 1\sigma$ interval.

Interestingly, we do find that on a few days the phase
distributions exhibit some clustering structure, implying a
marginal periodicity at the expected period. The phase
distributions for these days are shown in Figure~\ref{Fig6}. It could also be seen that there are even
multiple peaks in the profile on some days. Since each peak could
be produced by one emission site, this figure thus strongly
supports the multiple emission site interpretation. The low
significance of the periodicity on these days is due to the
simultaneous activeness of these sites.

We emphasize that this analysis is based on the two periods
detected on MJDs 59310 and 59347. It does not constitute an
independent discovery of periodicity on other days. Rather, it
shows clearly that on some non-detection days there are still
marginal periodical behaviors consistent with the same underlying
rotational clock.

\subsection*{Effects of multiple emission sites}

We have performed simulations to see how the number of
emission sites affects the periodicity of the observed bursts.
Assuming that FRBs originate from the magnetosphere of a rotating
magnetar, we generate a series of mock FRB events. In our
simulations, the rotation period of the magnetar is taken as $P =
1.707$ s. A number of $N$ bursts are generated in a time span of
$T_{\mathrm{obs}} = 2$ hr, which are assumed to originate from $m$
different emission sites randomly distributed over the rotation
phase. At each emission site, the bursts are clustered around the
central phase by following a Gaussian distribution. The
choice of the Gaussian standard deviation $\sigma$ could impact
the significance of the periodicity. A large $\sigma$ may smear
the periodicity completely. Based on the derived phase window
widths for MJDs 59310 and 59347, we adopted a moderate Gaussian
standard deviation of $\sigma=0.1$ periods in our simulations. The number
of bursts generated from each site was assumed to follow a multinomial distribution. There is a minimum phase gap of $g =
0.1$ periods between the central phases of adjacent emission
sites.

We have taken $N = 10,\ 20,\ 30,\ \ldots,\ 100$ and $m = 1,\ 2,\
3,\ \ldots,\ 10$ in our study, which leads to a total of 100
configurations. For each configuration, we generated $10^4$ simulated
TOA sequences. We then performed periodicity analysis on them and
calculated the corresponding average $\chi^2$ values. The results are
shown in Supplementary Figure~\ref{SFig4}. We see that as the
number of emission sites ($m$) increases, the periodicity becomes
undetectable. In fact, when there are more than two emission
sites, the periodicity will be essentially smeared and could not
be revealed even as many as 100 bursts are assumed. The non-detection of significant periodicity on days other than MJD 59310
and MJD 59347 in FRB 20201124A may be caused by the effects of
multiple emission sites.

\clearpage

\backmatter

\bmhead{Data availability} Most data used in this study are
available from the published literature. The burst data of FRB
20201124A are available from the following references:
\citep{hilmarsson2021Polarizationa,marthi2021Burst,xu2022fast,zhang2022FASTa}.
The dynamic spectra of the bursts are available in Ref.
\citep{wang2023Atlas}. Pulsar data used in this study are taken
from the ATNF Pulsar Catalogue \citep{manchester2005Australia}:
\url{https://www.atnf.csiro.au/research/pulsar/psrcat/}.

\begin{figure*}
\centering
\includegraphics[width=0.75\textwidth]{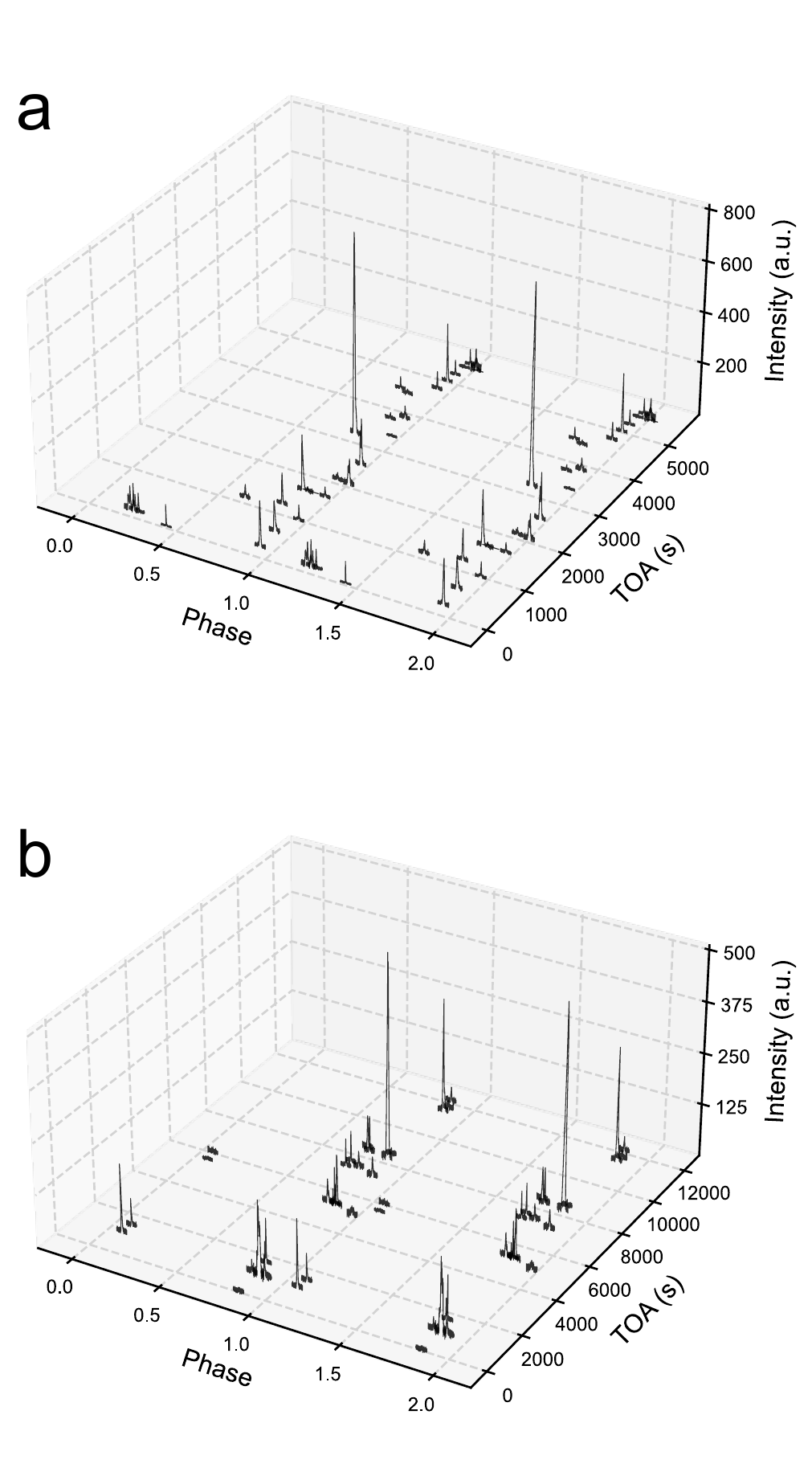}
\caption{{\bf The position of each burst in the phase space when folded
according to the refined period.} Bursts on MJDs 59310 and 59347
are illustrated in panels a and b, respectively. The X-axis is the folded phase, the
Y-axis shows the TOA, and the Z-axis illustrates the flux
density (in arbitrary units). For clarity of visualization, two
complete cycles are plotted. The mean TOA of bursts observed on
each day is taken as the time of the zero phase point. }
\label{Fig3}
\end{figure*}

\begin{figure*}
\centering
\includegraphics[width=1\textwidth]{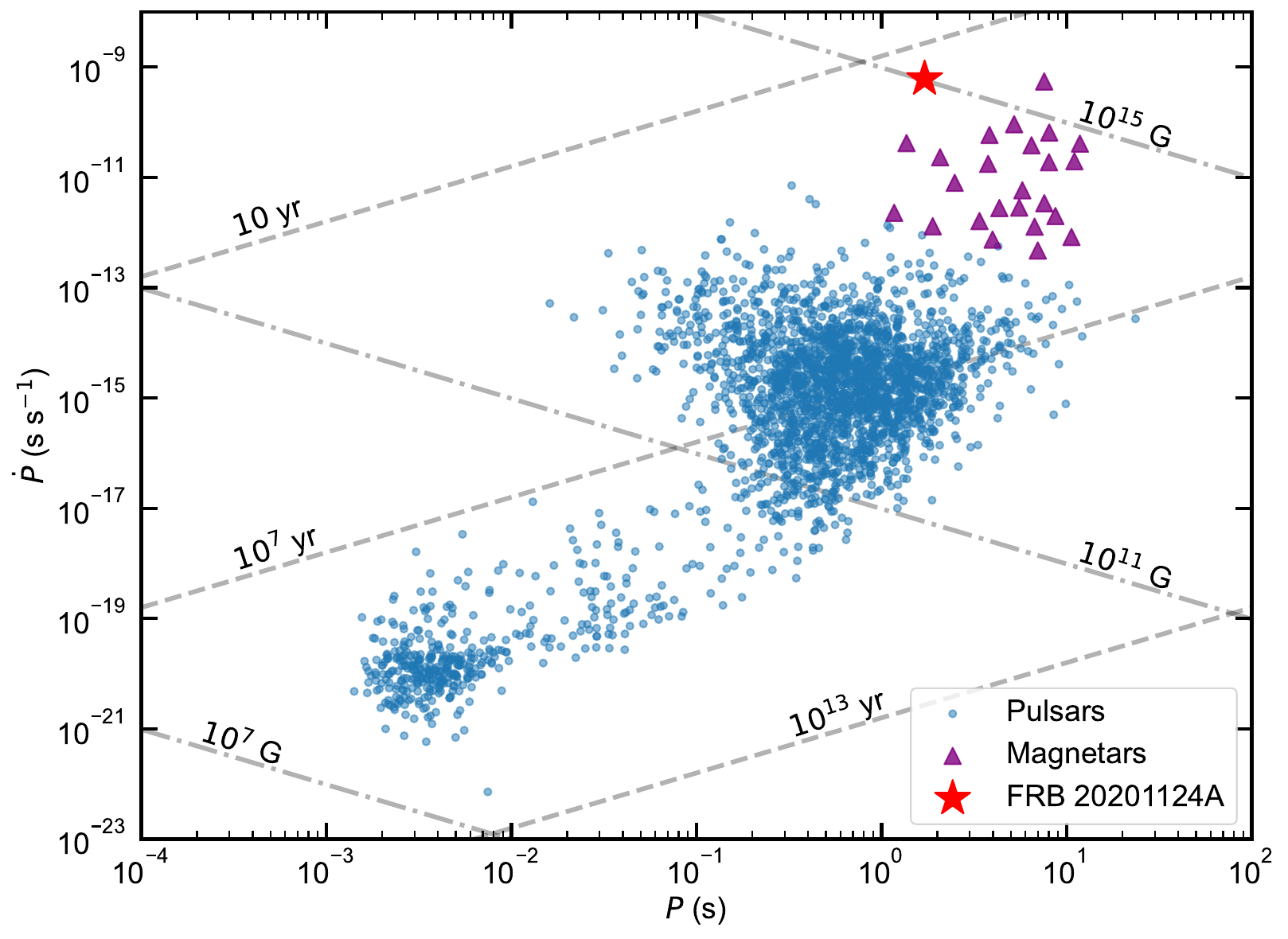}
\caption{{\bf The position of the young magnetar of FRB 20201124A as
compared with other pulsars on the $P$-$\dot{P}$ plane.} The
central engine of FRB 20201124A is marked by a pentagon. Normal
pulsars are indicated by dots and other magnetars are indicated by
triangles. The dashed-dotted lines denote the surface magnetic
field calculated under the magnetic dipole assumption. The dashed
lines denote the spin-down age.} \label{Fig4}
\end{figure*}

\begin{figure*}
\centering
\includegraphics[width=1\textwidth]{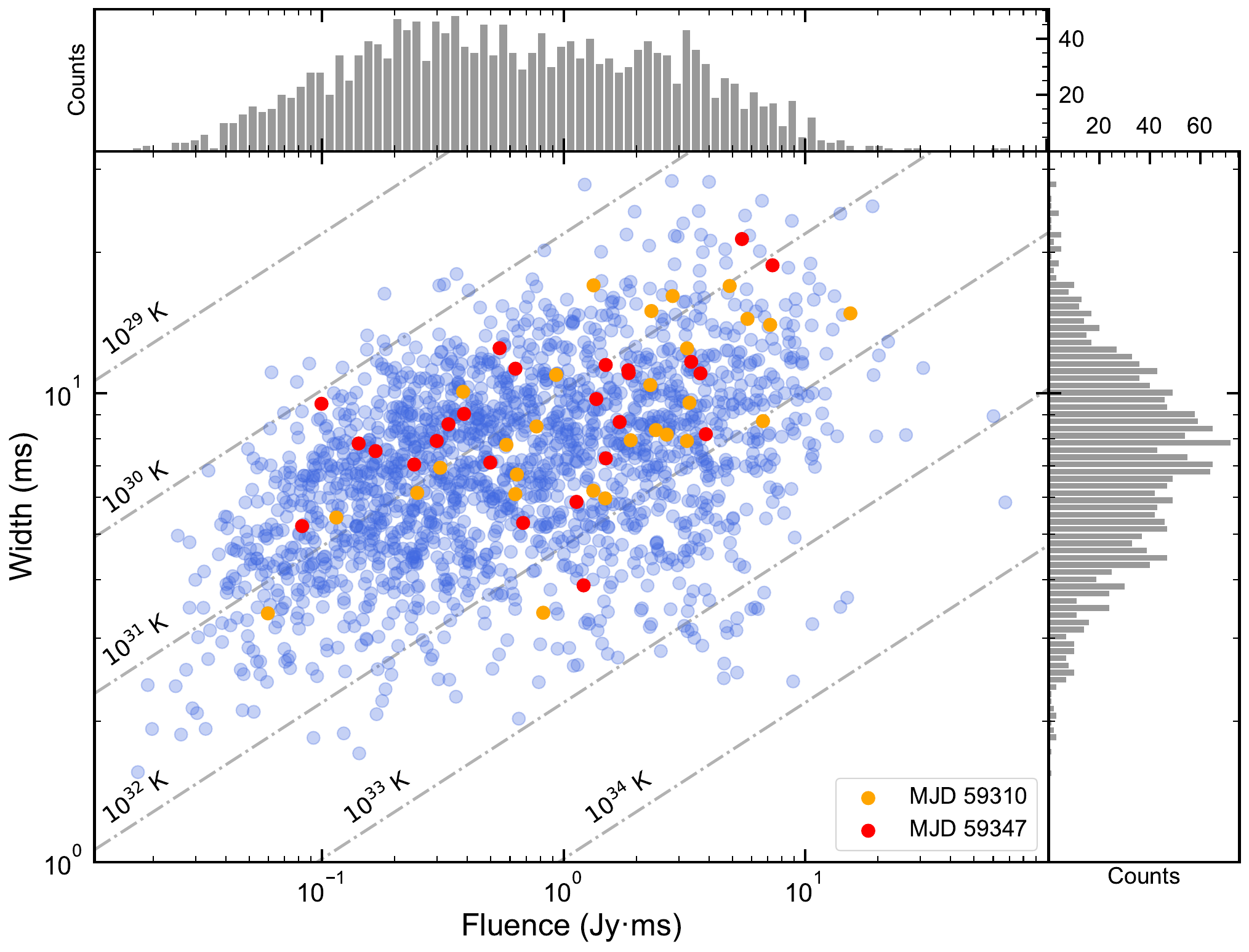}
\caption{{\bf The distribution of bursts from FRB 20201124A on the
width-fluence plane. All bursts are detected by FAST during the
first active episode.} The bursts on MJDs 59310 and 59347 are
marked by orange and red dots, respectively. The dashed-dotted
lines denote the brightness temperature. The observational data
are taken from Ref. \citep{xu2022fast}. } \label{Fig5}
\end{figure*}

\begin{figure*}
\centering
\includegraphics[width=1\textwidth]{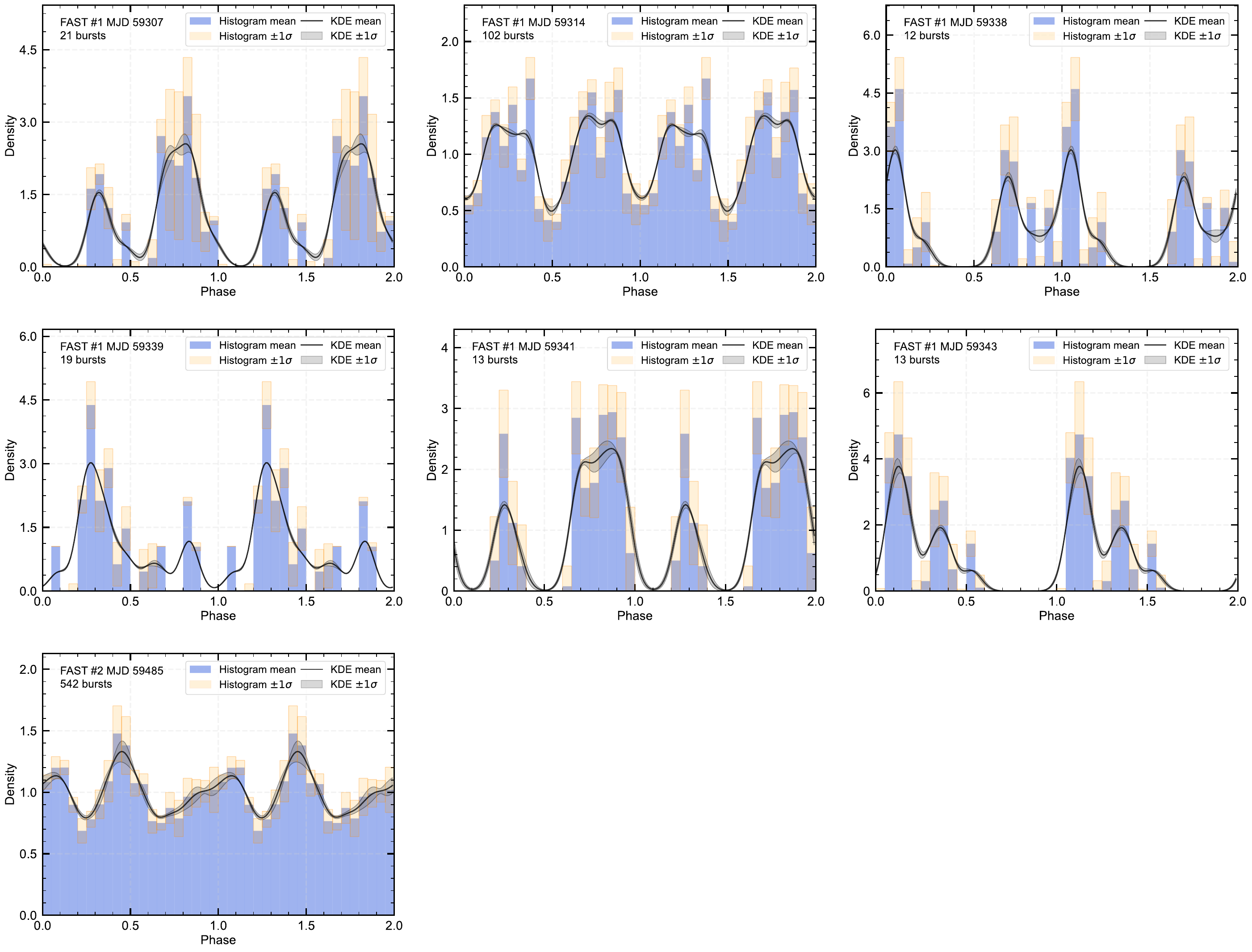}
\caption{{\bf Phase distributions of bursts on seven days when folded
according to the expected period.} For each day shown here, the
burst TOAs are folded using the period extrapolated from
the two detected periods on MJDs 59310 and 59347, assuming a
linear period evolution. The blue histograms show the mean phase
density, while the orange shaded regions mark the $\pm 1\sigma$ interval. The
black curves show the mean kernel-density estimate, and the gray
shaded bands indicate the corresponding $\pm 1\sigma$ interval. Two full
phase cycles are plotted for clarity. Although no significant
periodicity is revealed in blind searches on these days, this
figure shows clear clustering structures when the bursts are
folded using the extrapolated period. } 
\label{Fig6}
\end{figure*}

\begin{figure*}
\centering
\includegraphics[width=0.75\textwidth]{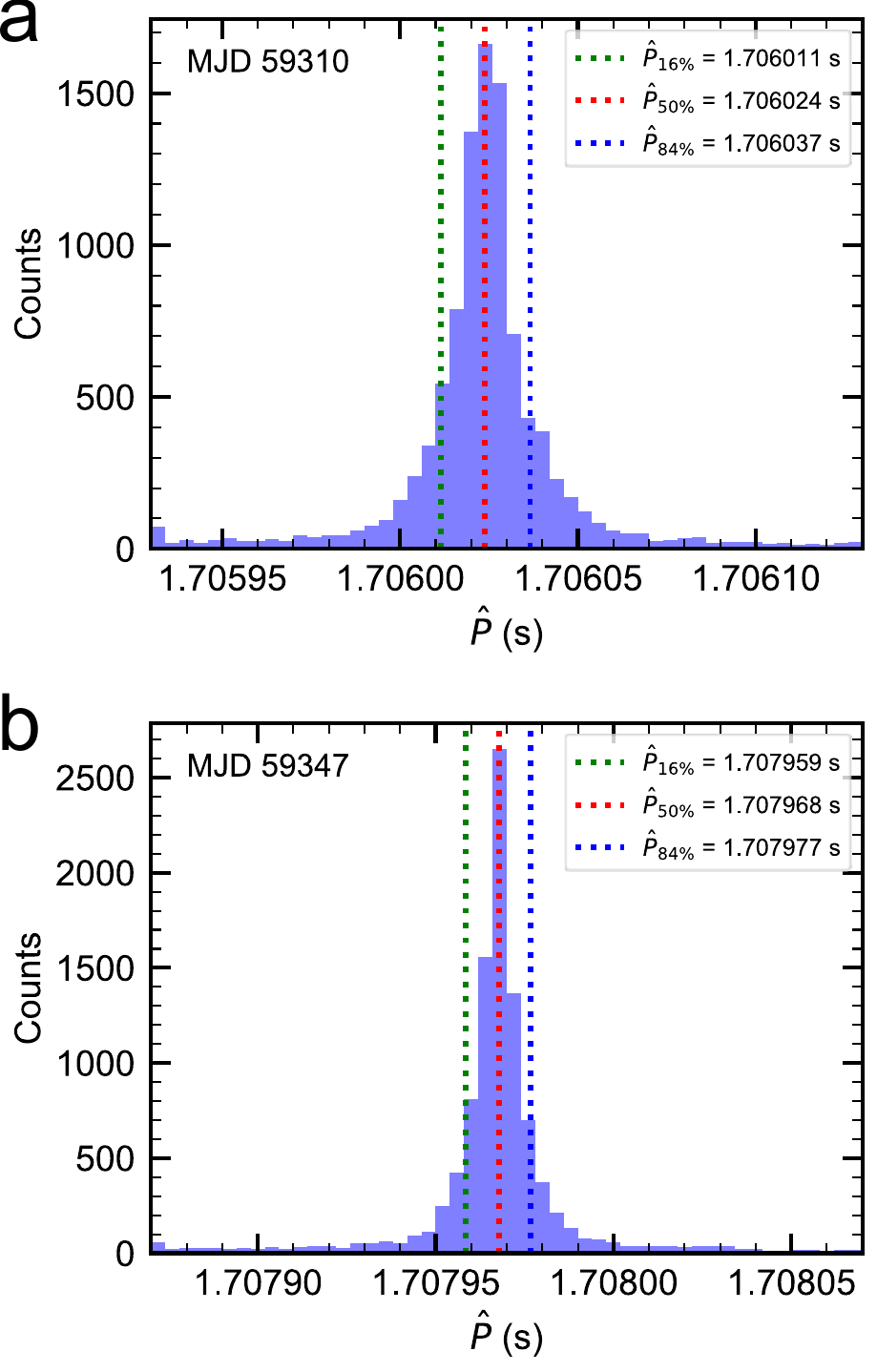}
\caption{{\bf Refining the period and estimating the error range by
using the parametric-bootstrap method.} The histograms show the
distributions of $\{\hat{P}\}$ of $10^{5}$ simulated parametric
bootstrap samples for MJD 59310 (panel a) and MJD 59347 (panel b). The
vertical dotted lines indicate the median $\hat{P}_{50\%}$ (red)
and the $\pm1\sigma$ bounds (green and blue).}
\label{Fig7}
\end{figure*}

\begin{figure*}
\centering
\includegraphics[width=1\textwidth]{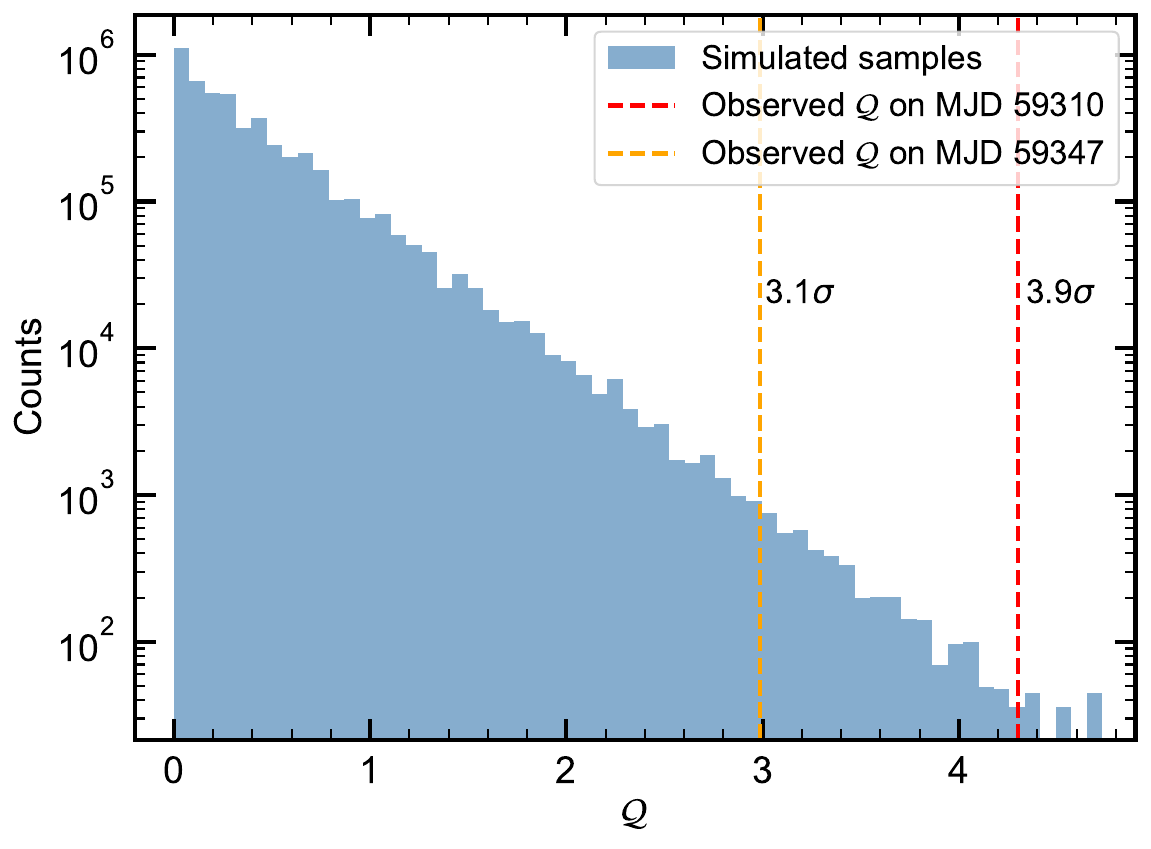}
\caption{{\bf Distribution of the pooled significance score
$\mathcal{Q}$ calculated by using simulated FRB samples.} For each
of the 51 single-day datasets, we generated $10^5$ mock samples.
All the mock samples were then used to calculate the significance
scores, which are shown as the histogram. The dashed vertical
lines indicate the observed $\mathcal{Q}$ values for the periods
on MJDs 59310 and 59347. They correspond to significance levels of
$3.9 \sigma$ and $3.1 \sigma$, respectively, suggesting that the
periodicity is significant.} 
\label{Fig8}
\end{figure*}

\begin{figure*}
\centering
\includegraphics[width=1\textwidth]{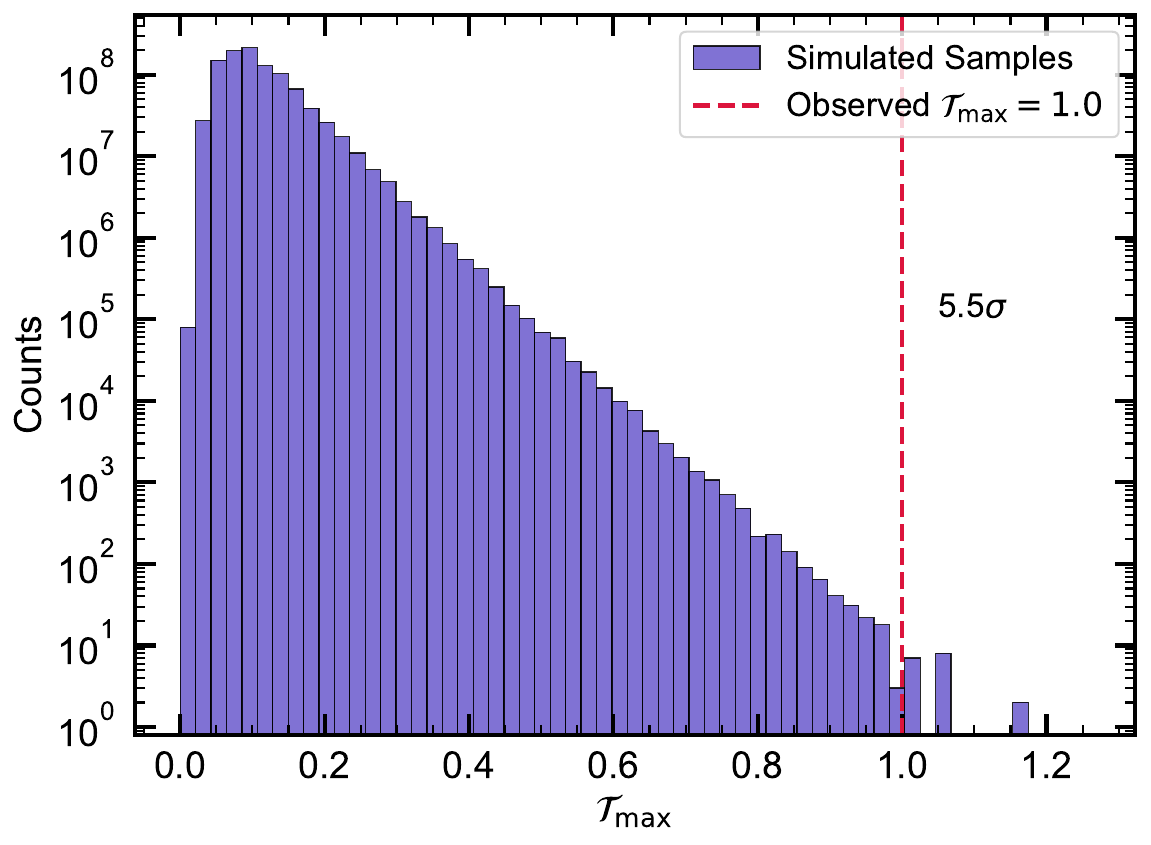}
\caption{{\bf Distribution of the global statistic $\mathcal{T}_{\max}$
for $10^9$ simulations.} To plot this figure, we first generated
$10^5$ mock samples for each of the 51 single-day datasets. One
mock sample is then randomly selected from the simulated samples
for each day. In this way, we get a mock dataset composed of 51
simulated samples, which was used to calculate the global
statistic $\mathcal{T}_{\max}$. The resampling process is repeated
for $10^9$ times to finally get the histogram in the figure. The
dashed vertical line marks the observed value,
$\mathcal{T}_{\max}=1$, corresponding to the MJD 59310--59347
pair. Only 20 out of the $10^9$ simulations satisfy
$\mathcal{T}_{\max}\geq1$, giving a global false-alarm probability
of $2\times10^{-8}$, equivalent to a significance level of
$5.5\sigma$.}
\label{Fig9}
\end{figure*}

\begin{figure*}
\centering
\includegraphics[width=1\textwidth]{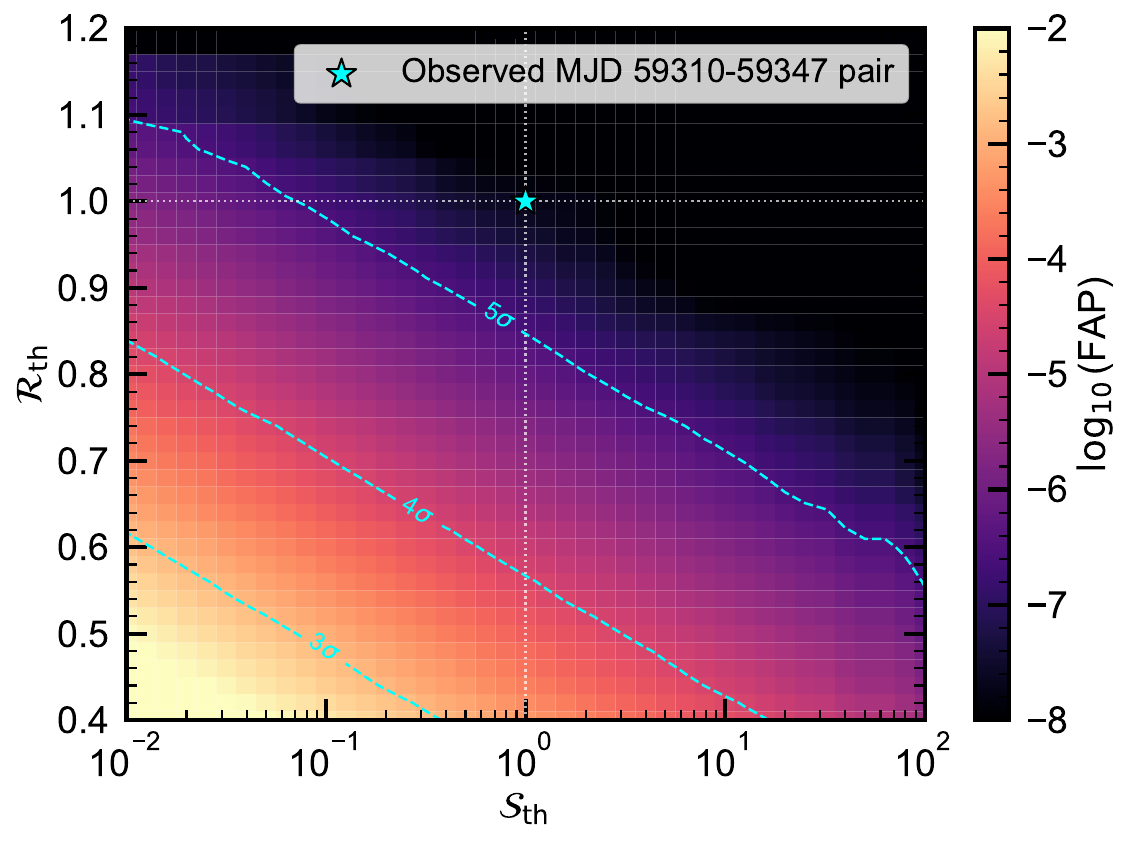}
\caption{{\bf Global false-alarm probability in the $(\mathcal{R}_{\rm
th},\mathcal{S}_{\rm th})$ plane.} For each set of
pair-significance threshold $\mathcal{R}_{\rm th}$ and
period-closeness threshold $\mathcal{S}_{\rm th}$, the
corresponding global false-alarm probability is computed. The
color scale shows $\log_{10}({\rm FAP})$. The cyan star marks the
observed MJD 59310--59347 pair, corresponding to
$(\mathcal{R}_{\rm th},\mathcal{S}_{\rm th})=(1,1)$. The dashed
contours indicate the significance levels of $3\sigma$, $4\sigma$,
and $5\sigma$. The observed point lies in the low-probability
region of the map, and the neighbouring grid points are also of
low probability, indicating that the inferred global significance
is robust.} 
\label{Fig10}
\end{figure*}

\bmhead{Code availability} \texttt{PSRCHIVE} (\url{https://psrchive.sourceforge.net}) was used to process the data presented in Figure~\ref{Fig3}. \texttt{PRESTO} (\url{https://github.com/scottransom/presto}) was used to search for bursts in our observations of FRB 20201124A. The codes used to perform period searches and generate figures can be found at \url{https://github.com/C-v-ke/searchforFRBperiodindepth}.

\bmhead{Acknowledgements} This study was supported by the National
Natural Science Foundation of China (Grant Nos. 12233002,
12273007, 12273113, 11963003, 12588202, 12573051), by the National SKA
Program of China (Nos. 2020SKA0120200, 2022SKA0130104),
%% and 2022SKA0120102,
%% CuiLang: 2022SKA0120102
by the National Key R\&D Program of China (Nos. 2021YFA0718500, 2022YFE0133700),
%% 2023YFE0102300),
%% LiuXiang: 2023YFE0102300
by the Guizhou Provincial Excellent Young Science and Technology Talent Program (No. YQK[2023]006), by the Guizhou Provincial Leading Talent Development Program (No.KJLYRC-[2026]007), by the Guizhou Provincial Basic Research Program (Natural Science) (No. QNA[2026]005),
and by the Major Science and
Technology Program of Xinjiang Uygur Autonomous Region
(No. 2022A03013-1).
%% by the CAS ``Light of West China'' Program (No. 2021-XBQNXZ-005),
%% and by Xinjiang Tianshan Talent Program.
%% CuiLang: 2021-XBQNXZ-005, Xinjiang Tianshan Talent Program
LC acknowledges the support from the Tianshan Talent Training Program (Grant No. 2023TSYCCX0099). YFH acknowledges the support from the Xinjiang Tianchi
Program. JJG acknowledges the support from the Youth Innovation Promotion Association CAS (2023331). PW acknowledges the support from the CAS Youth Interdisciplinary Team, the Youth Innovation Promotion Association CAS (2021055), and the Cultivation Project for FAST Scientific Payoff and Research Achievement of CAMS-CAS. CD1 (Chen Du) acknowledges the observations obtained with the Five-hundred-meter Aperture Spherical radio Telescope (FAST) under project PT2025\_0030.
\bmhead{Author contribution}
YFH proposed the research idea and designed the research plan. CD1 (Chen Du) reduced and analyzed the archival data, leading the period search calculations. CD1, YFH and LZ1 (Li Zhang) drafted the manuscript with suggestions from all co-authors. JJG and LZ1 participated in discussions on most of the scientific content and provided important suggestions for the illustrations. HXG, LZ1, CD2 (Chen Deng), LC, JL, PFJ, LZ2 (Liang Zhang), PW, CRH, XFD, FX, LL, ZCZ, and AK contributed comments and discussions that helped improve the research. All authors reviewed and commented on the manuscript.
\bmhead{Competing interests}
The authors declare no competing interests.

\clearpage

\clearpage

\clearpage

\clearpage

\clearpage

\clearpage

\clearpage

\clearpage

\bibliography{ms_nc}% common bib file
%% if required, the content of .bbl file can be included here once bbl is generated
%%\input sn-article.bbl

\clearpage
\begin{center}
    {\LARGE\bfseries Supplementary Information}
\end{center}
% \vspace{0.25cm}
% \hrule
\vspace{0.5cm}

\section*{Supplementary Discussion}
\subsection*{Comparison with the analyses of other groups}

After the first version of our manuscript was posted on arXiv
(arXiv:2503.12013v1), many researchers noticed our study and
examined the results. The periodicity on MJDs 59310 and 59347 was
confirmed by several researchers through independent analyses
using different methods \citep{suppZhang2025,suppShen2026}. However, Supplementary Ref.
\citep{suppGazith2026} claimed that the 1.7 s period is not
significant. Here we present a detailed comparison with their
analysis.

First, we notice that in Supplementary Ref. \citep{suppGazith2026}, the authors
admitted that they found some signature for the periodicity on MJD
59310. They wrote that ``MJD 59310 passes the bar of its 500
searches on sampled data, when analyzed using the $\chi^2$ test
and log-normal sampling strategy'', although they argued the
``result for MJD 59310 is not significant''. On MJD 59347, they
got $\chi^2 = 4.8$. Note that although this value is not so high,
it still points to some kind of periodicity which could be seen
from their Figure 3.

In the first arXiv version of our manuscript, we took the waiting
time threshold for identifying a cluster of bursts as 0.05 s (note
that the threshold has been taken as 0.5 s in a later version of our
study to be more consistent with other authors). A major
difference in Supplementary Ref. \citep{suppGazith2026} is that they used a much
larger waiting time threshold (0.4 s) as compared with our
previous value (0.05 s), which reduced the number of bursts by 3 on MJD 59347. Anyway, we have repeated their
calculations for MJDs 59310 and 59347 by following their method.
The results are shown in Supplementary Figure~\ref{SFig5}. We see
that for MJD 59310 we obtained a $\chi^2$ value of 5.22 at a
period of 1.7060169 s (consistent with their value of 5.2), which does support the existence of the
periodicity. For MJD 59347, we obtained a $\chi^2$ value of 4.84
at a period of 1.7079715 s (consistent with their value of 4.8),
still pointing to some kind of periodicity (although not so
significant).

In Supplementary Ref. \citep{suppGazith2026}, the authors took the first burst as
the representative event for a cluster of bursts with the waiting
time less than the threshold. This is not a natural selection. In
fact, it would be more reasonable to select the strongest burst as
the representative event. As illustrated in Supplementary
Figure~\ref{SFig6}, on MJD 59310 there are three clusters under
the 0.4 s waiting time threshold. In one cluster, the first burst
is not the brightest. On MJD 59347, there are two clusters under
the 0.4 s waiting time threshold, and in both clusters the first
burst is not the brightest. Consequently, retaining only the first
burst in each cluster introduces a systematic bias in the
representative TOA selection.

Bearing this in mind, we have re-done the analysis on MJDs 59310
and 59347, setting the waiting time threshold as 0.4 s but taking
the brightest burst to represent the cluster. The results are
shown in Supplementary Figure~\ref{SFig7}. We see that on MJD
59310, the $\chi^2$ value remains as high as 5.22 at a period of 1.7060169 s, which strongly supports the existence of the periodicity. On MJD 59347, the $\chi^2$ value is even higher, reaching 6.62 at a period of 1.707015 s, also supporting the periodicity.

We have further performed the $H$-test as a cross-check. In Supplementary Ref.
\citep{suppGazith2026}, the authors adopted a maximum harmonic number
of $m_{\max}=5$ when using the $H$-test, which is a bit too
coarse. For FRBs, the phase profile is complex and the $H$-test
power could be distributed among higher harmonics. We thus adopted
a maximum harmonic number of $m_{\max}=20$ to ensure the
sensitivity for complex phase profiles, which is also widely
adopted by other groups as a standard configuration of the
$H$-test \citep{suppdeJager2010}. Our $H$-test results for MJDs 59310
and 59347 over the 0.1--100 s range (with a step size of $10^{-5}$ s)
are shown in Supplementary Figure~\ref{SFig8}, using the
empirical $p$-value formula and applying a correction for the
number of trials ($\sim 10^7$). We see that on MJD 59310, an
obvious peak occurs at 1.70604 s, with a significance level of
3.4$\sigma$ (at 3.41207 s, a peak at the second harmonic is also
observed, corresponding to a significance level of 4.4$\sigma$).
It strongly supports the existence of periodicity on this day.
Similarly, on MJD 59347, the strongest peak is at 1.70797 s, with
a significance level of 5.2$\sigma$, also strongly supporting the
periodicity. Supplementary Table~\ref{STab2} summarizes the analytical
choices and resulting outcomes of Supplementary Ref. \citep{suppGazith2026}, as well
as our reproduction and our extended analysis.

We also note that, prior to our work, Supplementary Ref.
\citep{suppniu2022FAST} conducted dedicated searches for periodicity
in FRB 20201124A. It is worth emphasizing that the null result
reported in Supplementary Ref. \citep{suppniu2022FAST} is not in contradiction with
our period detection, but rather reflects differences in the
datasets involved and the methodologies employed. We summarize the
key distinctions between our analysis and that presented in Supplementary Ref. \citep{suppniu2022FAST} as
follows.

{\bf Datasets:} Supplementary Ref. \citep{suppniu2022FAST} focused primarily on the
second active episode of FRB 20201124A in September 2021,
analyzing the bursts detected over four days. In contrast, we
analyze a broader burst sample spanning 49 observing days, which
includes the first active episode from April to June 2021. The two
candidate days exhibiting the $\sim$1.7 s periodicity (MJDs 59310
and 59347) are in this earlier active episode. When applying our
search pipeline to the September 2021 data, we also find no
significant periodicity, which is consistent with the null result
of Supplementary Ref. \citep{suppniu2022FAST}.

{\bf Methodologies:} Supplementary Ref. \citep{suppniu2022FAST} employed multiple
approaches, including: (i) periodicity searches on the dedispersed
raw time series using \texttt{PRESTO} and \texttt{riptide}; (ii)
Lomb-Scargle periodicity searches on TOAs of bursts; (iii)
periodicity and acceleration search on TOAs of bursts with
$H$-test; (iv) periodicity search on TOAs of special bursts (such
as bursts with sign-changing circular polarization); (v) searches
for quasi-periodic structures within individual multi-component
bursts; and (vi) tests putative periods using waiting times. These
searches were mainly conducted over timescales ranging from
milliseconds to hundreds of seconds, depending on the method
itself. By contrast, our analysis focuses on searching for a
period associated with the spin of the central engine, using the
TOAs of bursts detected on each day separately to mitigate the
effects of $\dot{P}$. We use the phase-folding algorithm along
with the Pearson $\chi^2$ test as our primary analysis method,
while also using the $H$-test as a cross-check. Our searches were
conducted over the range from 0.1 s to 100 s.

{\bf Results:} No credible periodicity was identified in most of
the searches presented in Supplementary Ref. \citep{suppniu2022FAST}. Although some burst
substructures exhibited quasi-periodic behavior, these cannot be
regarded as the spin period of the magnetar. The authors therefore
emphasized that claims of spin periodicity in the multiple
components of FRBs should be treated with caution unless the
significance exceeds approximately $4\sigma$. To explain the
non-detection of periodicity, Supplementary Ref. \cite{suppniu2022FAST} suggested
that the burst activity may have a large duty cycle ($\gtrsim
0.5$). We detected a consistent $\sim$1.7 s periodicity on MJDs
59310 and 59347, from which we derived a period derivative of
$6.11(5)\times10^{-10}$ s s$^{-1}$, a value well within the
typical range for magnetars. The joint significance level of the
periodicity was estimated to be $5.5\sigma$ based on rigorous
end-to-end Monte Carlo simulations. We propose a scenario
involving multiple emission sites in the magnetosphere to account
for non-detections on other days.

\clearpage

\section*{Supplementary Figures}
\renewcommand{\figurename}{\textbf{Supplementary Fig.}}
\setcounter{figure}{0}
\renewcommand{\tablename}{\textbf{Supplementary Table}}
\setcounter{table}{0}

\begin{figure*}[htbp!]
\centering
\includegraphics[width=1\textwidth]{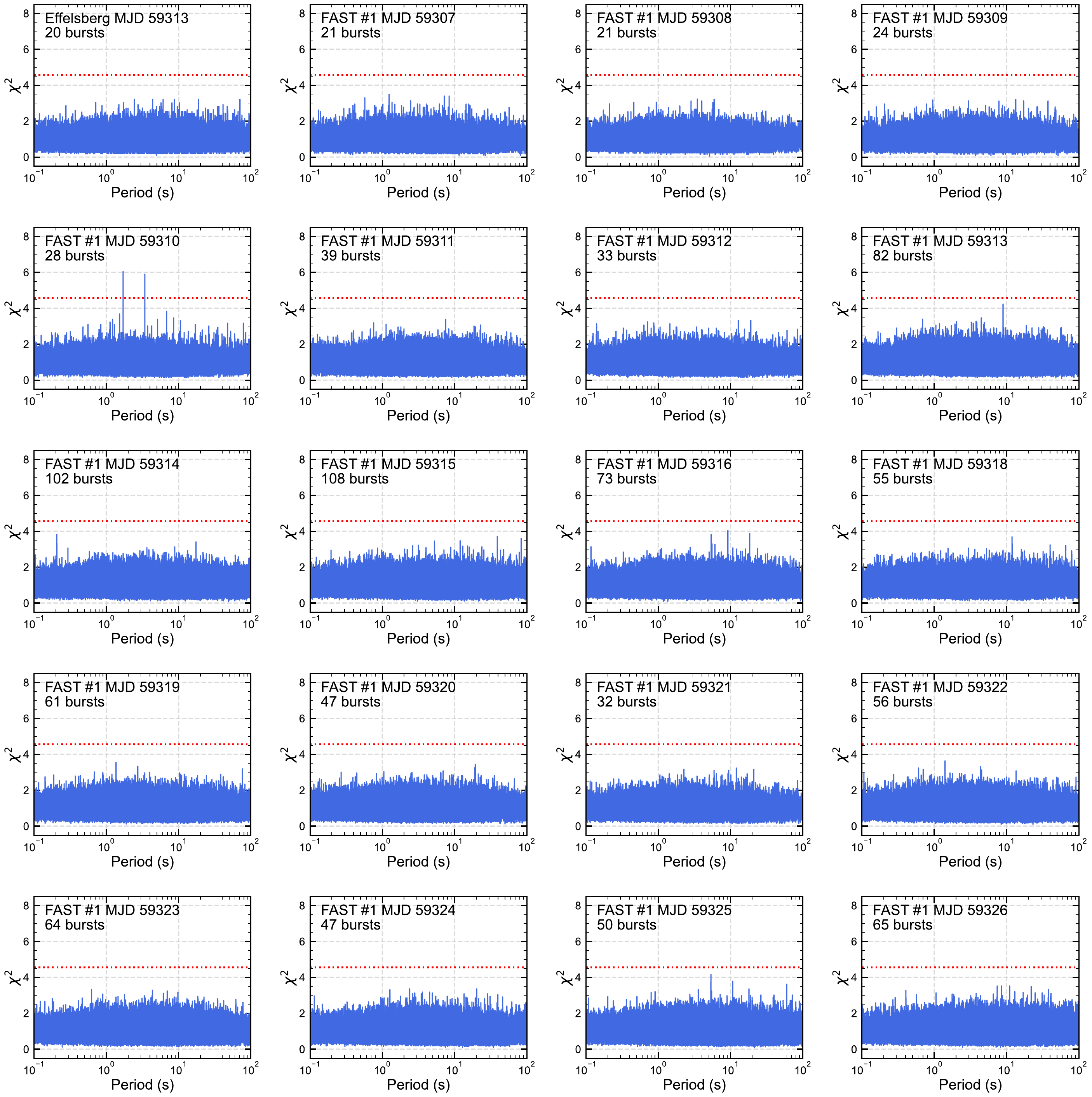}
\caption{{\bf $\chi^2$ periodograms of the bursts observed on each day for FRB 20201124A.}
The X-axis represents the trial period, and the Y-axis represents
the reduced $\chi^2$. The observing date and the telescope are
marked correspondingly in each panel. The number of bursts detected on each
day is also marked. The red dashed lines indicate a significance
level of $3\sigma$. To be continued on the next page.}\label{SFig1}
\end{figure*}

\begin{figure*}\ContinuedFloat
\centering
\includegraphics[width=1\textwidth]{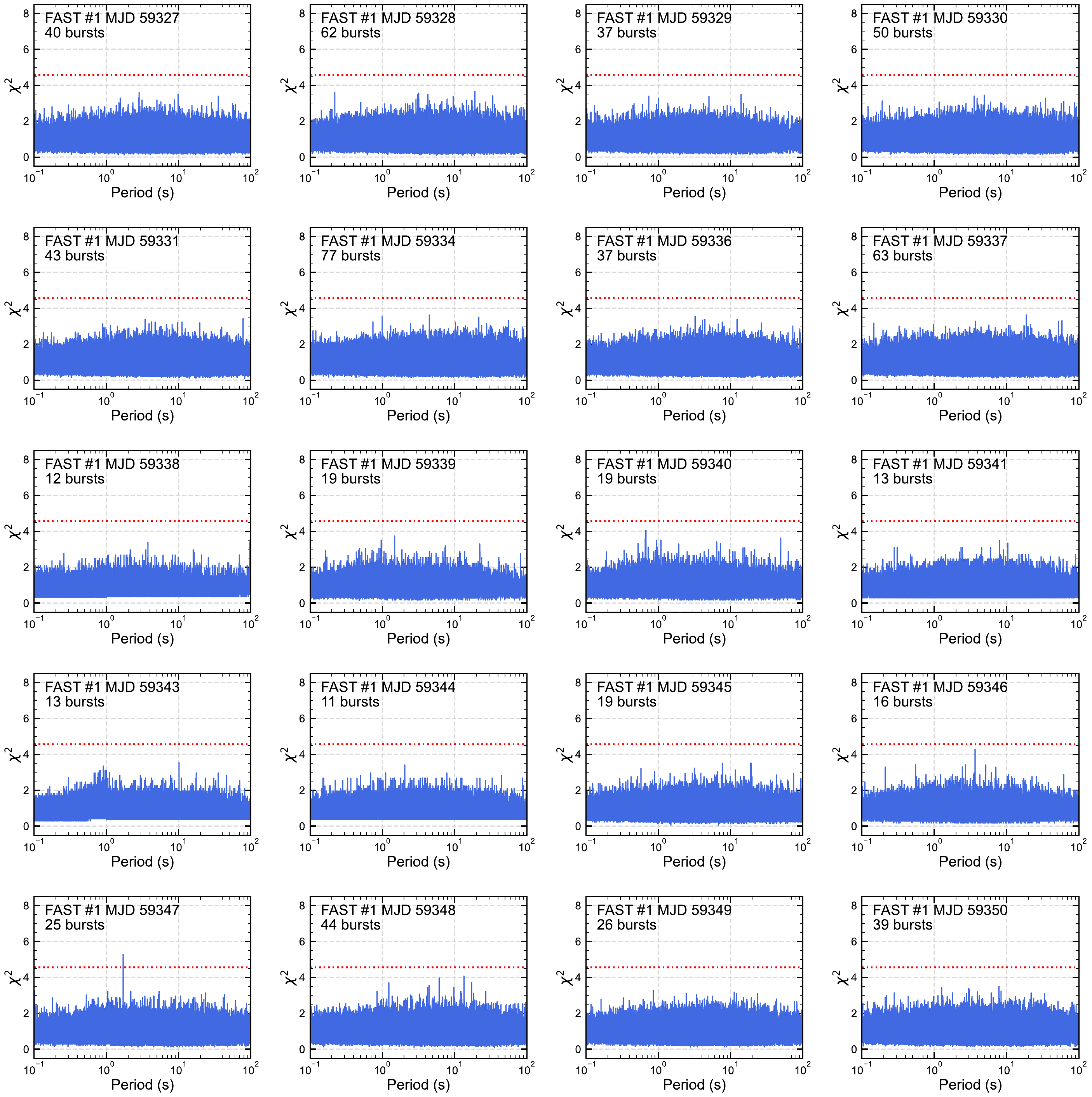}
\caption{{\bf $\chi^2$ periodograms of the bursts observed on each day for FRB 20201124A.}
The X-axis represents the trial period, and the Y-axis represents
the reduced $\chi^2$. The observing date and the telescope are
marked correspondingly in each panel. The number of bursts detected on each
day is also marked. The red dashed lines indicate a significance
level of $3\sigma$. To be continued on the next page.}
\end{figure*}

\begin{figure*}\ContinuedFloat
\centering
\includegraphics[width=1\textwidth]{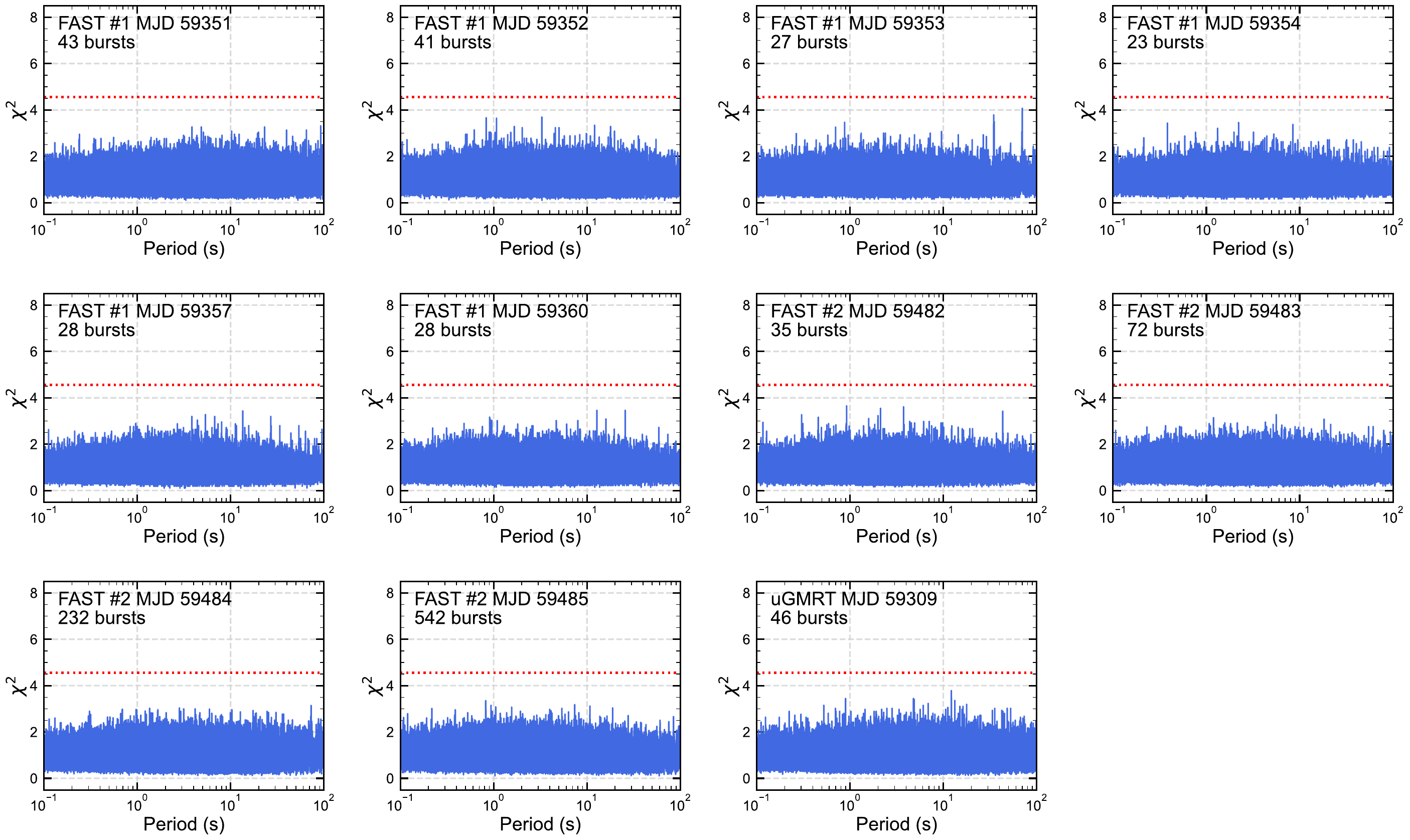}
\caption{{\bf $\chi^2$ periodograms of the bursts observed on each day for FRB 20201124A.}
The X-axis represents the trial period, and the Y-axis represents
the reduced $\chi^2$. The observing date and the telescope are
marked correspondingly in each panel. The number of bursts detected on each
day is also marked. The red dashed lines indicate a significance
level of $3\sigma$.}
\end{figure*}

\clearpage

\begin{figure*}
\centering
\includegraphics[width=1\textwidth]{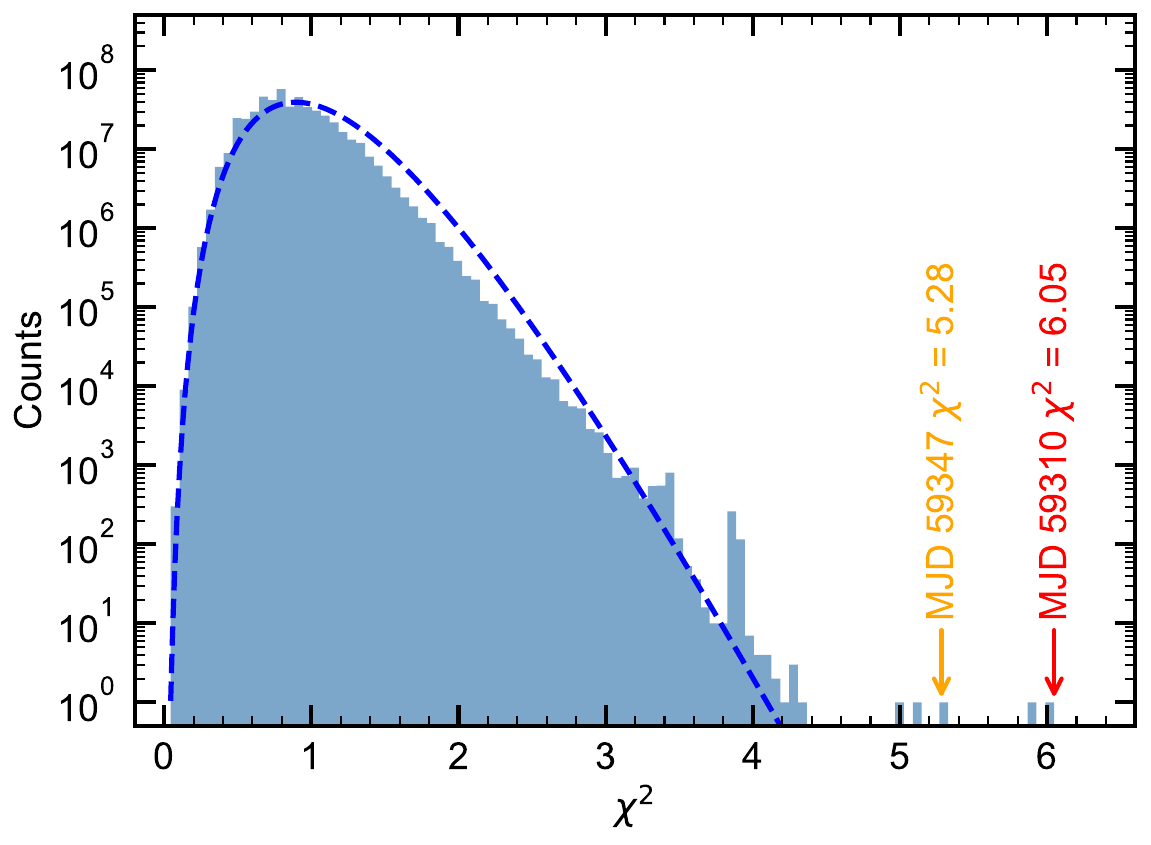}
\caption{{\bf Distribution of all our calculated $\chi^2$ values from
all the periodograms of the single-day blind search (blue
histogram).} The blue dashed line shows the theoretical $\chi^2$
curve, which is generally consistent with the histogram. The red
and orange arrows mark the peak $\chi^2$ values on MJDs 59310 and
59347, respectively, both of which lie in the extremely sparse
tail of the $\chi^2$ distribution. } \label{SFig2}
\end{figure*}

\begin{figure*}
\centering
\includegraphics[width=1\textwidth]{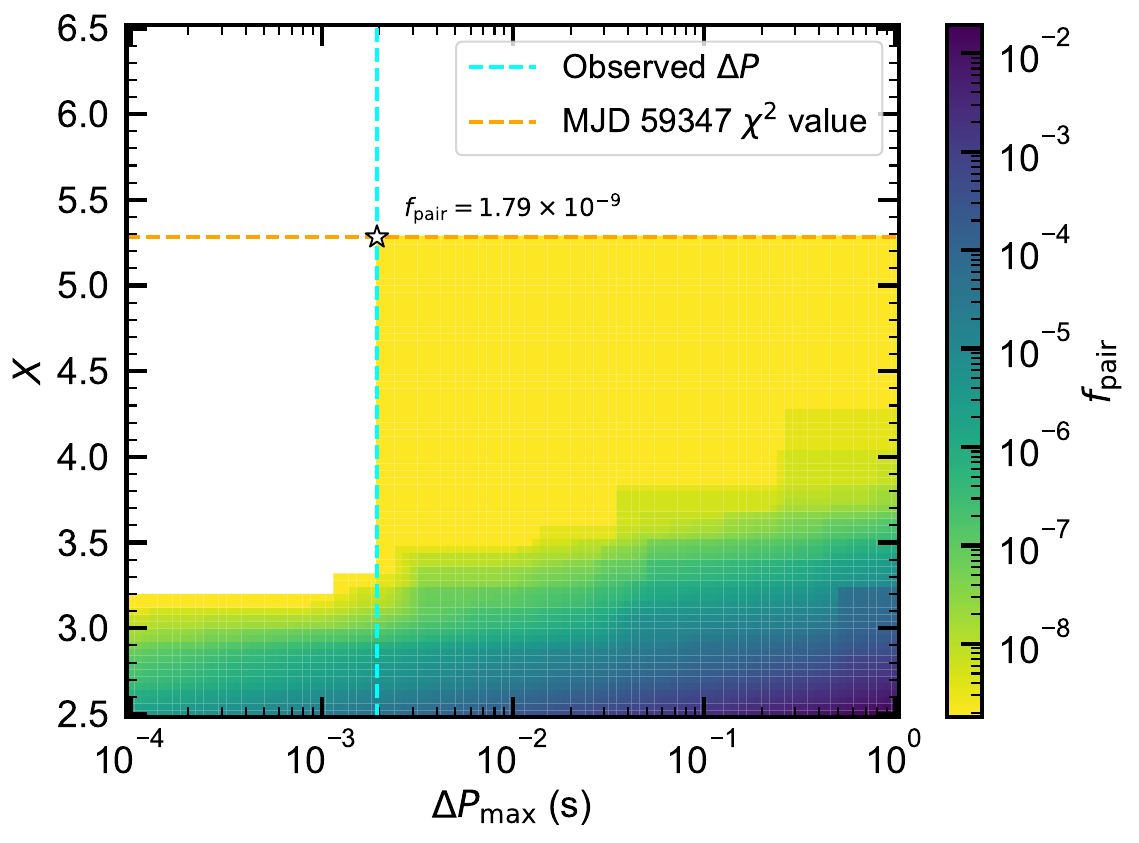}
\caption{{\bf Chance probability of finding close peak pairs across
periodograms of all single-day datasets.} For a given detection
threshold $X$ and maximum allowed period separation $\Delta
P_{\max}$, we counted the number of peak pairs satisfying both
criteria. The color scale shows the corresponding chance probability $f_{\rm pair}$, defined as the ratio of the number of
peak pairs satisfying both criteria to the total number of peak
pairs. The dashed cyan vertical line marks the observed period
separation between the two periods measured on MJDs 59310 and
59347, i.e., $\Delta P_{\rm obs}=0.00194$ s. The dashed orange
horizontal line marks the lower of the two peak powers, i.e.,
$\chi^2=5.28$ for MJD 59347. At the intersection, only one peak
pair, namely the MJD 59310--59347 pair, satisfies both criteria,
yielding $f_{\rm pair}=1.79\times10^{-9}$.} \label{SFig3}
\end{figure*}

\clearpage

\begin{figure*}
\centering
\includegraphics[width=1\textwidth]{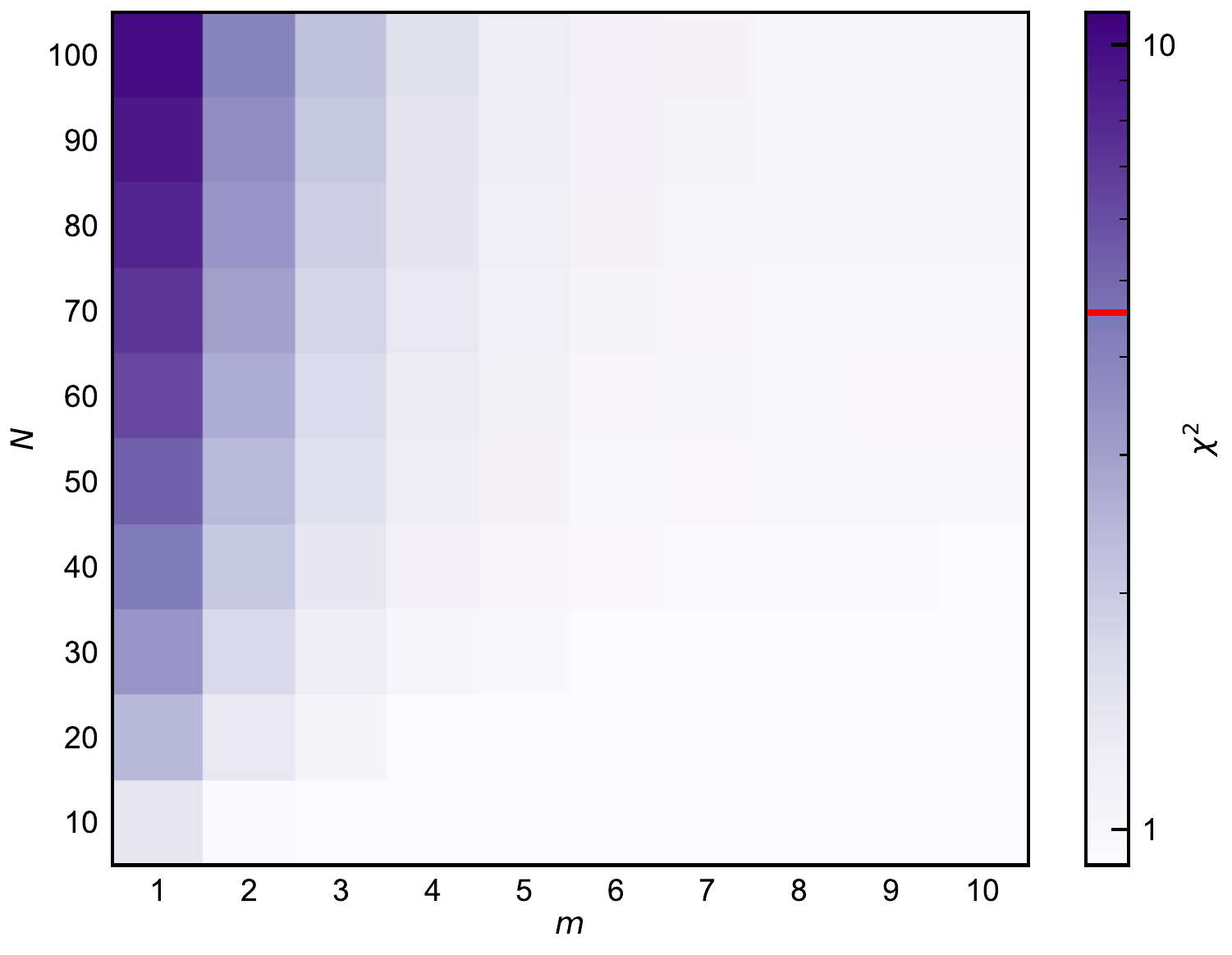}
\caption{{\bf Results of periodicity analysis for mock FRB samples.} In our simulations, a total number of $N$ bursts
are assumed to randomly originate from $m$ emission sites
inside the magnetosphere of a rotating magnetar. The
average $\chi^2$ value of the period searches is illustrated
for each configuration.
The red dash in the color bar indicates a significance
level of $3\sigma$.}
\label{SFig4}
\end{figure*}

\clearpage

\begin{figure*}
\centering
\includegraphics[width=0.9\textwidth]{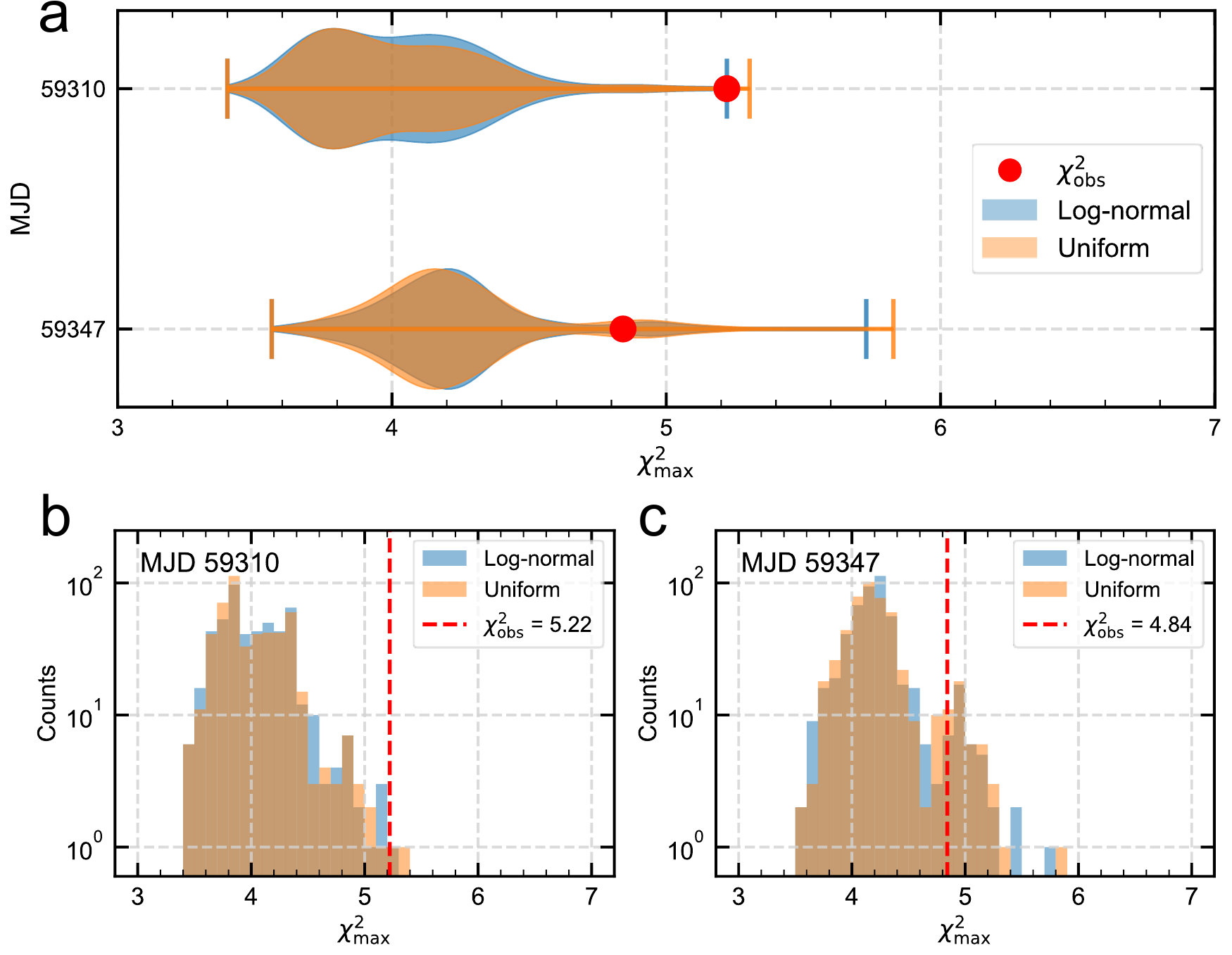}
\caption{{\bf Significance of the periodicity on MJDs 59310 and 59347.}
In this plot, a waiting time threshold of 0.4 s is adopted to
identify a cluster of bursts, and the first burst in each cluster
is selected as the representative event \citep{suppGazith2026}. Panel a: Violin plots summarizing the $\chi^2_{\rm max}$
distributions for MJDs 59310 and 59347. The red dots mark the
actual $\chi^2$ values ($\chi^2_{\rm obs}=5.22$ for MJD 59310 and
$\chi^2_{\rm obs}=4.84$ for MJD 59347). Similar to Supplementary Ref.
\citep{suppGazith2026}, we employ two null models: a ``log-normal''
model (blue) and a ``uniform'' model (orange). Panels b and c:
histograms of simulated $\chi^2_{\rm max}$ for the two days, with
the actual $\chi^2_{\rm obs}$ indicated by the red dashed vertical
lines. From this plot, we see that the periodicity on MJD 59310
is actually quite significant. The bursts on MJD 59347 also
exhibit periodical behavior to some extent.}
\label{SFig5}
\end{figure*}

\begin{figure*}
\centering
\includegraphics[width=0.9\textwidth]{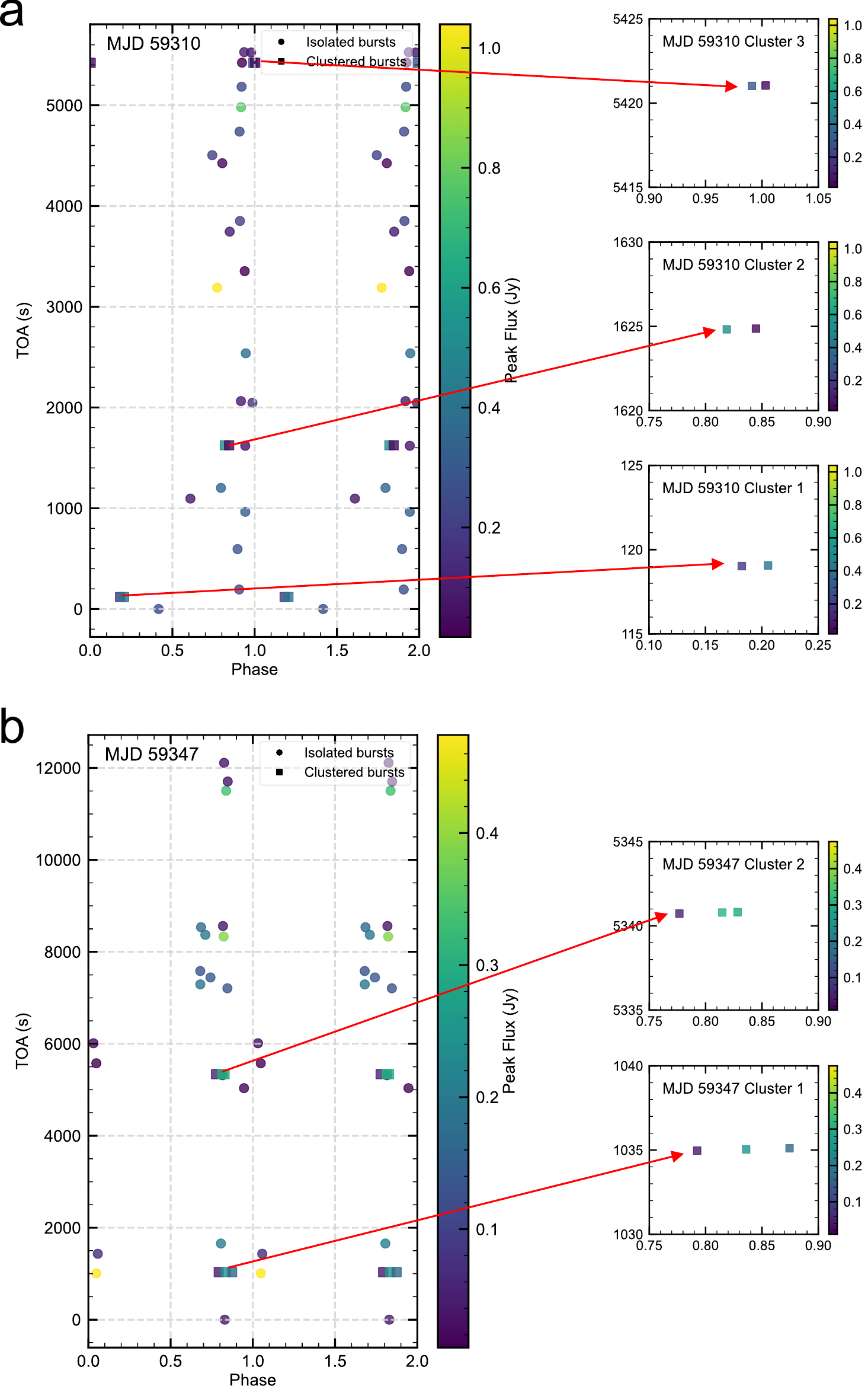}
\caption{{\bf Phase distribution of the bursts in each cycle on MJD
59310 (Panel a) and MJD 59347 (Panel b)}. The bursts are
folded by using the refined periods of 1.706024 s and 1.707968 s
on MJD 59310 and MJD 59347, respectively. The data points
representing the detected bursts are color-coded by their peak
flux densities. Bursts with a waiting time less than 0.4 s are
identified as a cluster and are shown as squares, while isolated
bursts are shown as filled circles. The smaller panels on the
right show zoomed views of the clustered bursts. On
MJD 59310, three clusters are identified. Note that in Cluster 1,
the second burst is brighter than the first one. On MJD 59347,
two clusters are identified. In both clusters, the first burst is
the weakest one among three.} \label{SFig6}
\end{figure*}

\clearpage

\begin{figure*}
\centering
\includegraphics[width=0.9\textwidth]{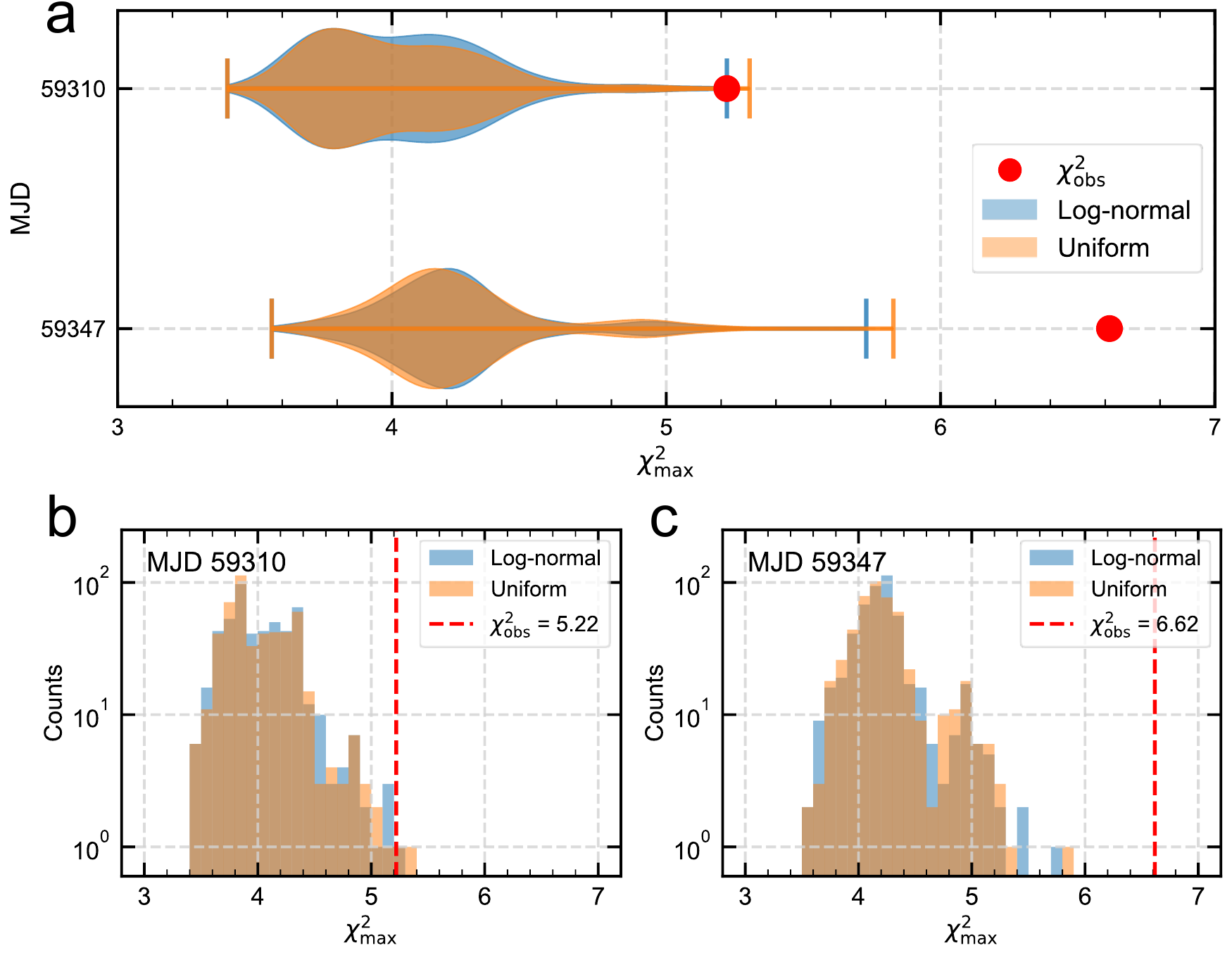}
\caption{{\bf Refined significance of the periodicity on MJDs 59310 and
59347.} This is a similar plot to Supplementary
Figure~\ref{SFig5}, except that the brightest burst in each
cluster is selected as the representative event for the cluster.
Panel a: Violin plots summarizing the $\chi^2_{\rm max}$
distributions for MJDs 59310 and 59347. The red dots mark the
actual $\chi^2$ values ($\chi^2_{\rm
obs}=5.22$ for MJD 59310 and $\chi^2_{\rm obs}=6.62$ for MJD
59347). Similar to Supplementary Ref. \citep{suppGazith2026}, we employ two null
models: a ``log-normal'' model (blue) and a ``uniform'' model
(orange). Panels b and c: histograms of simulated $\chi^2_{\rm max}$ for the two days, with the actual $\chi^2_{\rm obs}$
indicated by the red dashed vertical lines. From this plot, we see that on MJD 59310, the $\chi^2$ value remains high, supporting the existence of the periodicity. On MJD 59347, the $\chi^2$ value is even higher, reaching 6.62, which further provides strong evidence for the periodicity.} 
\label{SFig7}
\end{figure*}

\begin{figure*}
\centering
\includegraphics[width=1\textwidth]{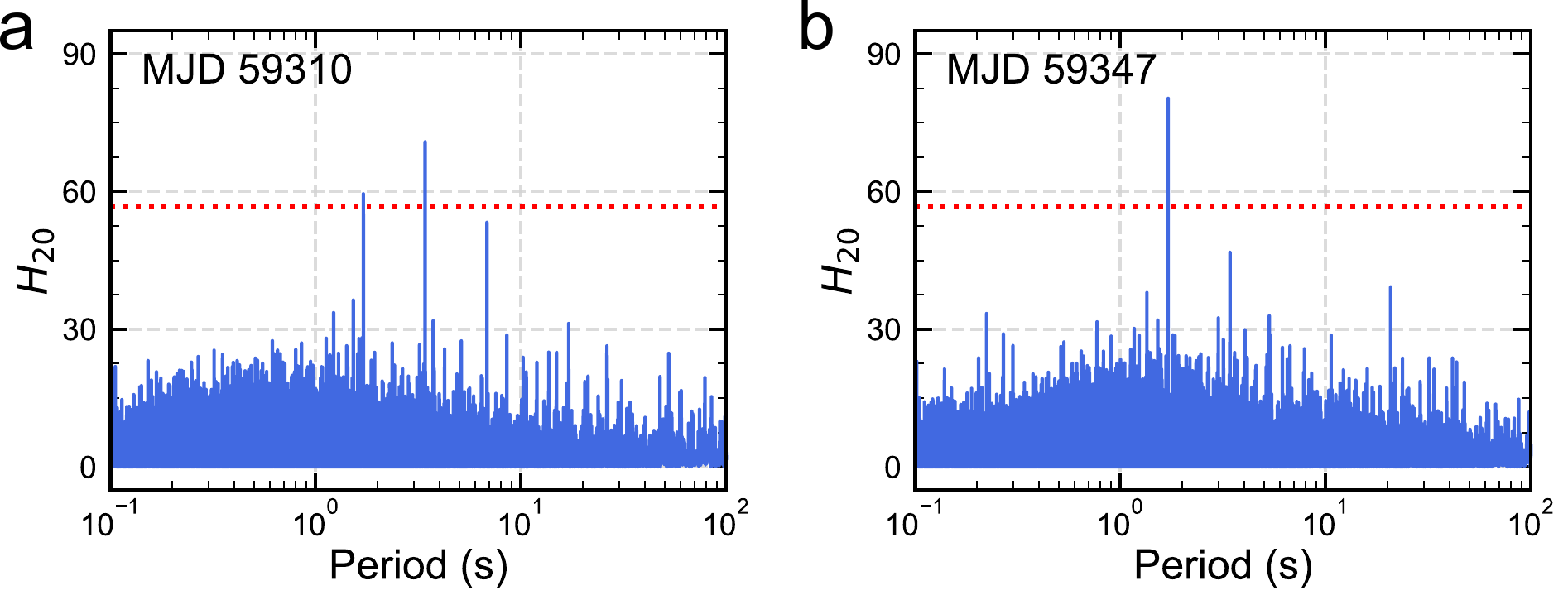}
\caption{{\bf The $H$-test periodograms for the bursts on MJD 59310
(panel a) and MJD 59347 (panel b).} The dotted horizontal line
corresponds to a significance level of $3\sigma$ derived from the
empirical $p$-value formula. On MJD 59310, a clear periodicity at
1.70604 s could be seen, with a significance level of 3.4$\sigma$.
Its second harmonic could also be observed at 3.41207 s, with a
significance level of 4.4$\sigma$. On MJD 59347, a clear
periodicity is observed at 1.70797 s, with a significance level of
5.2$\sigma$.} \label{SFig8}
\end{figure*}

\clearpage

\section*{Supplementary Tables}

\captionof{table}{{\bf Log of all the continuous observing sessions used in our study on FRB 20201124A.}  Column 1 is the start time of the session. Column 2 is the end time of the session. Column 3
presents the duration of the session in seconds. Column 4 is
the number of bursts detected in the session.}
\label{STab1}
\footnotesize \setlength{\tabcolsep}{5pt} \noindent
\begin{minipage}[t]{0.45\textwidth}
\begin{tabular}{ccccc}
            \toprule
            Start & End & Duration & Bursts \\
            (MJD)   & (MJD)   & (s) &  counts \\
            \midrule
              59307.33299 & 59307.41632 & 7200 & 21 \\
              59308.33187 & 59308.41520 & 7200 & 21 \\
              59309.34566 & 59309.42900 & 7200 & 24 \\
              59309.52163 & 59309.68830 & 14400 & 46 \\
              59310.32474 & 59310.40808 & 7200 & 28 \\
              59311.32533 & 59311.40866 & 7200 & 39 \\
              59312.26832 & 59312.35166 & 7200 & 33 \\
              59313.32452 & 59313.40786 & 7200 & 82 \\
              59313.74119 & 59313.90786 & 14400 & 20 \\
              59314.32126 & 59314.40460 & 7200 & 102 \\
              59315.31825 & 59315.40159 & 7200 & 108 \\
              59316.31770 & 59316.40104 & 7200 & 73 \\
              59318.31039 & 59318.31441 & 348 & 2 \\
              59318.31763 & 59318.32397 & 548 & 5 \\
              59318.34860 & 59318.39355 & 3884 & 48 \\
              59319.30351 & 59319.38685 & 7200 & 61 \\
              59320.22669 & 59320.30656 & 6900 & 47 \\
              59321.22314 & 59321.28565 & 5401 & 32 \\
              59322.29597 & 59322.37931 & 7200 & 56 \\
              59323.29590 & 59323.37924 & 7200 & 64 \\
              59324.33054 & 59324.33961 & 783 & 7 \\
              59324.34563 & 59324.41792 & 6245 & 40 \\
              59325.24367 & 59325.26451 & 1801 & 15 \\
              59325.28533 & 59325.30618 & 1801 & 6 \\
              59325.32700 & 59325.34784 & 1801 & 12 \\
              59325.36866 & 59325.38950 & 1801 & 17 \\
              59326.22276 & 59326.24361 & 1801 & 12 \\
              59326.26443 & 59326.28527 & 1801 & 12 \\
              59326.30609 & 59326.31936 & 1147 & 10 \\
              59326.32733 & 59326.34817 & 1801 & 14 \\
              59326.34977 & 59326.37061 & 1801 & 17 \\
              59327.21575 & 59327.23659 & 1801 & 5 \\
              59327.25741 & 59327.27826 & 1801 & 18 \\
              59327.29908 & 59327.31992 & 1801 & 10 \\
              59327.34074 & 59327.35429 & 1171 & 7 \\
              59328.26429 & 59328.28513 & 1801 & 10 \\
              59328.30595 & 59328.32679 & 1801 & 15 \\
              59328.34761 & 59328.36846 & 1801 & 19 \\
              59328.38928 & 59328.41012 & 1801 & 18 \\
              59329.37913 & 59329.44163 & 5401 & 37 \\
              59330.24018 & 59330.26102 & 1801 & 13 \\
              59330.28151 & 59330.30235 & 1801 & 14 \\
              59330.32317 & 59330.34402 & 1801 & 19 \\
              59330.36484 & 59330.38568 & 1801 & 4 \\
              59331.26237 & 59331.28321 & 1801 & 17 \\
              59331.30366 & 59331.32451 & 1801 & 11 \\
              59331.34533 & 59331.36617 & 1801 & 8 \\
              59331.38699 & 59331.40783 & 1801 & 7 \\
              59334.22570 & 59334.24654 & 1801 & 17 \\
              59334.26736 & 59334.28820 & 1801 & 16 \\
              59334.30902 & 59334.32987 & 1801 & 27 \\
              59334.35763 & 59334.37848 & 1801 & 17 \\
              59336.21168 & 59336.23253 & 1801 & 8 \\
              59336.24640 & 59336.26725 & 1801 & 13 \\
              59336.30890 & 59336.32975 & 1801 & 10 \\
              59336.35057 & 59336.37141 & 1801 & 6 \\
              59337.20468 & 59337.22553 & 1801 & 14 \\
              59337.24635 & 59337.26719 & 1801 & 16 \\
              59337.28801 & 59337.30885 & 1801 & 13 \\
              59337.32967 & 59337.35052 & 1801 & 20 \\
              59338.27754 & 59338.29838 & 1801 & 7 \\
              59338.37822 & 59338.39907 & 1801 & 5 \\
            \bottomrule
\end{tabular}
\end{minipage}
\hspace{0.07\textwidth}
\begin{minipage}[t]{0.45\textwidth}
\begin{tabular}{ccccc}
            \toprule
            Start & End & Duration & Bursts \\
            (MJD)   & (MJD)   & (s) &   counts   \\
            \midrule
              59339.25665 & 59339.27749 & 1801 & 15 \\
              59339.31220 & 59339.33304 & 1801 & 4 \\
              59340.18368 & 59340.20452 & 1801 & 5 \\
              59340.21840 & 59340.23924 & 1801 & 4 \\
              59340.25312 & 59340.27396 & 1801 & 6 \\
              59340.28784 & 59340.30869 & 1801 & 4 \\
              59341.20132 & 59341.22217 & 1801 & 3 \\
              59341.24265 & 59341.26350 & 1801 & 2 \\
              59341.28432 & 59341.30516 & 1801 & 5 \\
              59341.32598 & 59341.34682 & 1801 & 3 \\
              59343.19047 & 59343.21131 & 1801 & 6 \\
              59343.23213 & 59343.25298 & 1801 & 1 \\
              59343.27380 & 59343.29464 & 1801 & 5 \\
              59343.31546 & 59343.33631 & 1801 & 1 \\
              59344.22306 & 59344.24390 & 1801 & 2 \\
              59344.26472 & 59344.28556 & 1801 & 3 \\
              59344.30638 & 59344.32723 & 1801 & 5 \\
              59344.34805 & 59344.36889 & 1801 & 1 \\
              59345.22856 & 59345.24941 & 1801 & 2 \\
              59345.27023 & 59345.29107 & 1801 & 5 \\
              59345.31189 & 59345.33274 & 1801 & 6 \\
              59345.35356 & 59345.37440 & 1801 & 6 \\
              59346.27504 & 59346.29589 & 1801 & 5 \\
              59346.30282 & 59346.32366 & 1801 & 3 \\
              59346.33059 & 59346.35160 & 1815 & 5 \\
              59346.35837 & 59346.37922 & 1801 & 3 \\
              59347.16042 & 59347.18126 & 1801 & 7 \\
              59347.21045 & 59347.23130 & 1801 & 7 \\
              59347.24375 & 59347.26459 & 1801 & 8 \\
              59347.28541 & 59347.30626 & 1801 & 3 \\
              59348.18676 & 59348.20761 & 1801 & 10 \\
              59348.22843 & 59348.24927 & 1801 & 12 \\
              59348.27009 & 59348.29094 & 1801 & 12 \\
              59348.32148 & 59348.34232 & 1801 & 10 \\
              59349.16242 & 59349.18326 & 1801 & 5 \\
              59349.20408 & 59349.22493 & 1801 & 9 \\
              59349.24575 & 59349.26659 & 1801 & 5 \\
              59349.29497 & 59349.31581 & 1801 & 7 \\
              59350.16238 & 59350.18322 & 1801 & 6 \\
              59350.20404 & 59350.22489 & 1801 & 8 \\
              59350.25332 & 59350.27417 & 1801 & 18 \\
              59350.30260 & 59350.32345 & 1801 & 7 \\
              59351.16234 & 59351.18318 & 1801 & 9 \\
              59351.19359 & 59351.21443 & 1801 & 11 \\
              59351.22483 & 59351.24568 & 1801 & 9 \\
              59351.25608 & 59351.27693 & 1801 & 14 \\
              59352.24563 & 59352.26648 & 1801 & 8 \\
              59352.27688 & 59352.29772 & 1801 & 11 \\
              59352.30813 & 59352.32897 & 1801 & 11 \\
              59352.33938 & 59352.36022 & 1801 & 11 \\
              59353.23171 & 59353.25255 & 1801 & 14 \\
              59353.27337 & 59353.29422 & 1801 & 13 \\
              59354.13793 & 59354.15877 & 1801 & 13 \\
              59354.17265 & 59354.19349 & 1801 & 10 \\
              59357.13089 & 59357.15173 & 1801 & 13 \\
              59357.16561 & 59357.18645 & 1801 & 15 \\
              59360.12386 & 59360.14471 & 1801 & 12 \\
              59360.15859 & 59360.17943 & 1801 & 16 \\
              59482.94316 & 59482.98483 & 3600 & 35 \\
              59483.86261 & 59483.90427 & 3600 & 72 \\
              59484.81400 & 59484.85566 & 3600 & 232 \\
              59485.78275 & 59485.82441 & 3600 & 542 \\
            \bottomrule
\end{tabular}
\end{minipage}

\clearpage

\begin{table}[htbp]
\centering
 \caption{{\bf Comparison of our extended analysis result with that of Supplementary Ref. \citep{suppGazith2026}.}}
\label{STab2}
\renewcommand{\arraystretch}{1.5}
\footnotesize \setlength{\tabcolsep}{4.6pt} \noindent
\begin{tabular}{lcccccc}
\toprule
Analysis
& \begin{tabular}[c]{@{}c@{}}De-clustering\\threshold\end{tabular}
& \begin{tabular}[c]{@{}c@{}}Representative\\burst\end{tabular}
& \begin{tabular}[c]{@{}c@{}}$m_{\rm max}$\\in $H$-test\end{tabular}
& \begin{tabular}[c]{@{}c@{}}Candidate\\MJD\end{tabular}
& $\chi^2$
& $H$ \\
\midrule
In Supplementary Ref. \citep{suppGazith2026} & 0.4\,s & first & 5 &
\begin{tabular}[c]{@{}c@{}}59310\\59347\end{tabular} &
\begin{tabular}[c]{@{}c@{}}5.2\\ 4.8\end{tabular}
& \begin{tabular}[c]{@{}c@{}}53\\ 38\end{tabular} \\
\midrule
\begin{tabular}[c]{@{}l@{}}Our reproduction of\\Supplementary Ref. \citep{suppGazith2026}\end{tabular}
& 0.4\,s
& first
& 5
& \begin{tabular}[c]{@{}c@{}}59310\\59347\end{tabular}
& \begin{tabular}[c]{@{}c@{}}5.22\\ 4.84\end{tabular}
& \begin{tabular}[c]{@{}c@{}}52.4\\ 36.1\end{tabular} \\
\midrule
\begin{tabular}[c]{@{}l@{}}Our extended analysis using\\method of Supplementary Ref. \citep{suppGazith2026}\end{tabular}
& 0.4\,s
& brightest
& 20
& \begin{tabular}[c]{@{}c@{}}59310\\59347\end{tabular}
& \begin{tabular}[c]{@{}c@{}}5.22\\ 6.62\end{tabular}
& \begin{tabular}[c]{@{}c@{}}59.5\\ 80.3\end{tabular} \\
\bottomrule
\end{tabular}

\vspace{2mm}
\begin{minipage}{1\linewidth}
\small \textit{Note.} To present a direct comparison with the
results of Supplementary Ref. \citep{suppGazith2026}, we have re-analyzed the bursts
on MJDs 59310 and 59347 by using the method similar to that of
Supplementary Ref. \citep{suppGazith2026}. The first block here lists the analytical
choices and resulting outcomes reported in Supplementary Ref.
\citep{suppGazith2026}. The second block illustrates our reproduction
of Supplementary Ref. \citep{suppGazith2026} setup, using the same de-clustering
threshold, the same representative burst choice (the first burst
in each cluster), and the same maximum harmonic number in the
$H$-test. The close agreement between the outcomes in the first
and second blocks demonstrates that we did not misunderstand the
method of Supplementary Ref. \citep{suppGazith2026}. Note that the significance of
the periodicity on MJD 59310 is already quite high. The third
block shows our extended analysis based on the same de-clustering
threshold, in which the representative burst is changed to the
brightest burst in each cluster and the $H$-test is evaluated with
$m_{\rm max} = 20$. We see that while the periodicity on MJD 59310
is still significant, the periodicity significance on MJD 59347 is
much improved. It indicates that the periodicity should be
authentic.
\end{minipage}
\end{table}

\renewcommand{\refname}{Supplementary References}

\end{document}